\documentclass[11pt]{article}
\usepackage[margin=1.07in]{geometry}
\usepackage{amsmath,amssymb,amsthm,mathtools,mathrsfs,booktabs,microtype}
\usepackage[authoryear,round]{natbib}
\allowdisplaybreaks[1]
\newtheorem{theorem}{Theorem}[section]
\newtheorem{proposition}[theorem]{Proposition}

\newtheorem{corollary}[theorem]{Corollary}
\newtheorem{remark}[theorem]{Remark}
\theoremstyle{definition}
\newtheorem{definition}[theorem]{Definition}
\theoremstyle{plain}

\newcommand{\E}{\mathbb E}

\newcommand{\D}{\mathscr D}

\begin{document}

\title{Optimal Information Disclosure in Classic Auctions Revisited: \\
Endogenous Information Acquisition and Costly Verification}
\author{Guoqiang Tian}
\date{First draft: July 2026; Revised: October 27, 2026}

\maketitle

\begin{abstract}
We revisit the influential distribution-independent optimal disclosure result of Bergemann, Heumann, Morris, Sorokin, and Winter (2022), whose optimal disclosure policy reveals low bidder values and pools high values at a critical quantile depending only on the number of bidders, and show that this policy is no longer generally revenue-maximizing even for the familiar uniform and binary valuation distributions, once bidders can endogenously acquire private information through costly verification. The underlying reason reflects a fundamental principle of incentive-based mechanism design, rooted in Hurwicz's analysis of structural invariance and echoed in the Lucas critique: economic agents respond strategically to changes in the information and incentives they face. This is the essence of \emph{trust, but verify}. A change in disclosure policy may leave the underlying economic primitives unchanged while altering bidders' verification decisions and the resulting equilibrium information structure. We establish existence and uniqueness of the symmetric Bayesian Nash verification equilibrium, characterize equilibrium-implementable disclosure policies, and derive an exact revenue decomposition and a general characterization of global optimality. We further develop optimality conditions and structural results that provide a systematic foundation for analyzing optimal disclosure with endogenous information acquisition.
\end{abstract}


\pagebreak
\setcounter{tocdepth}{2}
\tableofcontents

\section{Introduction}

How much information should a seller disclose to bidders in an auction? Providing more information can help bidders determine what an object is worth to them and improve the allocation of the object. But better-informed bidders may also obtain greater information rents, reducing the seller's revenue. Bergemann, Heumann, Morris, Sorokin, and Winter (2022, henceforth BHMSW) study this trade-off in a classic second-price auction in which bidders initially do not know their own valuations and the seller chooses what information they receive. They establish a striking result: the revenue-maximizing disclosure policy fully reveals low values while pooling high values. Under this low-value-revelation/high-value-pooling policy, the cutoff is determined by the number of bidders in quantile space, independently of the underlying distribution of valuations. Although developed for second-price auctions, BHMSW explain that their benchmark result also applies to classical revenue-equivalent auction formats, including first-price and all-pay auctions, under the relevant independent private-value assumptions.

The strength of this result lies in its simple disclosure structure and distribution-independent critical quantile. It is therefore natural to ask whether the same policy remains optimal when the bidders themselves can respond to the seller's disclosure by producing information. This is not a change in the auction's allocation rule or in the prior distribution of values. It is a change in what bidders can do after receiving the information the seller chooses to provide. As we show, that additional margin is enough to overturn the generality of the benchmark disclosure conclusion.

In many auctions, however, bidders do not rely exclusively on information supplied by the seller. Consider auctions of valuable artworks, antiques, and other collectibles. Sellers provide descriptions and information about the objects offered for sale, but prospective buyers may also conduct their own investigations to determine what those objects are worth to them. Such behavior follows a familiar principle: trust, but verify. Even when the seller's information is reliable, it need not be sufficient for a bidder to assess her own valuation. Further investigation can provide valuable information, but requires time, effort, or money. The bidder must decide whether obtaining that information is worth its cost.

For a buyer considering a painting or an antique, the question is not simply whether to believe a catalogue entry. Information can be credible and still incomplete for deciding how much to bid. A buyer may examine provenance records, commission an independent appraisal, inspect the object, or seek evidence about authenticity and condition. Doing so takes resources; choosing not to do so leaves the buyer exposed to uncertainty. The familiar injunction to ``trust, but verify'' therefore has a direct strategic meaning in an auction: what a bidder learns privately depends on the information already disclosed, the cost of further investigation, and the bids she expects to face.

This possibility changes the seller's disclosure problem. Pooling high values may increase revenue by limiting the information rents bidders obtain from the seller's disclosure. Yet the uncertainty left by pooling can also encourage bidders to acquire additional information privately. The seller therefore faces two related considerations: how disclosure directly affects competition and revenue, and how it changes bidders' incentives to verify their values. Each bidder faces a corresponding trade-off between the expected benefit of making a better-informed bidding decision and the cost of verification. Moreover, the value of verification depends on how well informed competing bidders are. The seller's disclosure policy and bidders' information-acquisition decisions are consequently linked through equilibrium behavior.

The principle of ``trust, but verify'' reflects a deeper lesson of incentive-based economic analysis: agents do not passively accept the information supplied by a designer, but respond to the incentives created by the designer's policy. This observation echoes \citet{Hurwicz1962Structural}'s fundamental insight that structural relationships must be understood relative to the changes under consideration, and the related concern underlying the \citet{Lucas1976} critique that policy changes may alter agents' behavioral decision rules. In the present setting, a change in disclosure policy leaves the underlying valuation environment and verification technology unchanged, but changes bidders' incentives to acquire information and hence the equilibrium informational environment in which the auction operates. The seller cannot therefore evaluate alternative disclosure policies while treating bidders' information-acquisition decisions as invariant. Information disclosure is not merely a choice of what information to reveal; it is also an intervention that changes the incentives to learn.

This interaction has a central implication. The distribution-independent optimal disclosure result established by Bergemann et al. (2022)---revealing low values while pooling high values at a critical quantile depending only on the number of bidders---does not remain generally revenue maximizing when verification is an endogenous choice. We establish the failure of that recommendation in two of the most elementary and familiar valuation environments in auction theory: uniform and binary values. The significance of these findings is not that a specially tailored distribution creates an exception. Rather, the benchmark policy already loses its generality in settings that normally serve as the starting point for auction analysis. The original BHMSW theorem remains valid under its own assumptions; what does not survive is the extension of its optimal disclosure conclusion to a world in which bidders can privately verify their values.

The economic reason is the seller's interaction with bidders' information production. Revealing information can improve bidding decisions while raising information rents; withholding it may limit those rents while inducing bidders to learn for themselves. A bidder, in turn, weighs the expected advantage of better information against the cost of verification, taking into account how informed her competitors are likely to become. Thus neither the seller's disclosure choice nor a bidder's verification choice can be assessed in isolation. The relevant object is the Bayesian Nash equilibrium induced by each disclosure policy, and the seller compares revenues across those equilibria.

We study this interaction by extending the independent private-value environment of BHMSW. The seller continues to choose symmetric, independent private information experiments about bidders' original valuations, and the object is sold through a second-price auction. After receiving the seller's signal, however, each bidder may privately undertake costly verification to learn more about her own valuation. Verification is a hidden action, and its outcome provides additional private information without changing the underlying valuation. We focus on a simple information-production technology in which the bidder chooses an effort that determines the probability of learning her valuation perfectly. The seller anticipates the equilibrium verification decisions when choosing the initial information structure.

The distinction between information disclosed by the seller and information acquired by bidders is important. In the original disclosure problem, the seller can choose the distribution of bidders' posterior expected valuations, subject to Bayes plausibility. Once verification is possible, the same distribution of posterior means may induce different information-acquisition incentives. A bidder's willingness to verify depends not only on her expected valuation but also on the uncertainty that remains about her actual value. Thus, the seller must take account of the full posterior beliefs generated by disclosure and the subsequent information choices of bidders.

The failure of the benchmark policy also creates a constructive problem: if the seller cannot treat the final information in bidders' hands as something she directly controls, how should optimal disclosure be characterized? Our analysis proceeds from the bidders' Bayesian Nash equilibrium in private verification to the seller's feasible equilibrium outcomes, and then to general revenue and optimality conditions. This positive theory complements the main economic finding rather than merely providing a way to calculate the examples.

Our first contribution, developed in Section~3, is to characterize the equilibrium of this information-acquisition game. A bidder's gain from verification depends on the distribution of competing bids, which in turn depends on their verification decisions. This strategic feedback leads to a fixed-point problem that must be resolved before evaluating the seller's disclosure policy. We show that the symmetric verification equilibrium admits a convex-potential representation, which establishes existence and uniqueness under strictly convex verification costs. The associated equilibrium Jacobian has a positive-semidefinite Gram structure, providing a tractable characterization of equilibrium interactions and comparative statics. We extend the existence and uniqueness results from finite signal structures to arbitrary posterior distributions using fixed-point and potential arguments. The resulting equilibrium characterization also establishes continuity of the final-bid distribution with respect to the seller's information experiment, a key step in proving the existence of an optimal disclosure policy over the unrestricted experiment space.

Our second contribution, developed in Section~5, concerns optimal disclosure for general distributions of bidder values. We characterize which final-bid distributions can arise from a Bayes-plausible private disclosure experiment together with the bidders' equilibrium verification choices. The characterization applies to the unrestricted space of private information experiments rather than a prespecified family of disclosure policies. Using convex separation and concavification, we obtain necessary and sufficient conditions for equilibrium implementability, together with a posterior-price dual representation. These results provide an infinite-dimensional characterization of feasible disclosure outcomes without requiring concavity of the seller's revenue function or imposing a particular form on the optimal experiment.

This general characterization addresses a difficulty created by endogenous information acquisition. Although the seller chooses the initial information structure, she does not directly choose the final distribution of bids. That distribution results from the combination of disclosure and bidders' privately chosen verification efforts. Consequently, standard information-design arguments based only on distributions of posterior means are no longer sufficient. Our characterization identifies the additional equilibrium restrictions that a final-bid distribution must satisfy. It provides a general description of feasible disclosure outcomes without imposing a particular structure on the seller's optimal experiment.

In Sections~5 and 6, to study the structure of optimal disclosure more directly, we derive the revenue effects of changes in the seller's information policy, allowing verification efforts to adjust in equilibrium. Using the implicit-function theorem, we separate the direct effect of disclosure on bidding from the indirect effect operating through endogenous verification. We also develop a rare-signal splitting test that introduces new private signals while preserving Bayes plausibility. The resulting directional condition accounts for both equilibrium effort adjustments and changes in the distribution of bids, including moving bid atoms. This provides a rigorous test of whether a proposed disclosure policy can be improved and plays a central role in establishing the failure of the benchmark disclosure structure.

Our third contribution, established in Section~4, is a sharp result on that question. We first characterize verification equilibrium under the benchmark policy for general distributions of bidder values. In this case, bidders whose values are fully revealed have no reason to verify, while those receiving the pooled high-value signal face a nontrivial information-acquisition decision. The equilibrium therefore reduces to a one-dimensional problem. For three bidders with uniformly distributed valuations, we obtain closed-form expressions for verification effort and seller revenue, allowing us to study how endogenous verification changes the optimal pooling cutoff.

The analysis in Section~4 shows that the benchmark disclosure conclusion of BHMSW does not generally survive endogenous information acquisition. Consider three bidders with independent uniform valuations and quadratic verification costs. We first identify the revenue-maximizing policy within the entire class of policies that reveal lower values and pool higher values. We then construct an admissible refinement of its high-value pool and account for the verification equilibrium induced by the refined information structure. The refinement yields strictly higher expected revenue than the optimal policy within that benchmark class. The comparison is established analytically by an exact polynomial root count and rational revenue bounds, rather than inferred from a numerical comparison of selected disclosure policies.

The economic mechanism behind this result is the interaction between pooling and private verification. Pooling high values can benefit the seller when bidders do not acquire additional information. But the same pooling leaves uncertainty that bidders may find worthwhile to resolve. Refining the high-value pool changes not only what bidders initially know but also how much additional information they choose to acquire. The seller can therefore gain from a disclosure policy that would not be optimal if bidders' information-acquisition decisions were held fixed. The failure of low-value revelation and high-value pooling arises from this equilibrium response, not from any change in the auction format or in bidders' original valuation distributions.

Our findings distinguish the value of partial disclosure from the particular structure of optimal partial disclosure. The result of BHMSW remains valid in their original environment, which is recovered when additional information acquisition is unavailable. In our extended environment, full disclosure need not be optimal either. What changes is the revenue-maximizing way of withholding and revealing information. The counterexample establishes that no policy in that benchmark class is optimal in the specified environment, although it does not identify an explicit optimal experiment over the unrestricted disclosure space. The general equilibrium and implementability results provide a framework for studying that broader problem.

Our analysis relates to two strands of research. The first studies information disclosure and Bayesian persuasion, where a designer chooses an information structure to influence subsequent decisions. BHMSW apply this approach to classic auctions and establish the benchmark result that serves as our point of departure. The second studies endogenous information acquisition, including bidders' incentives to learn before participating in auctions. Our contribution connects these approaches: the seller chooses an information structure while anticipating that bidders will respond by privately acquiring additional information. This interaction changes both the set of equilibrium bidding outcomes and the structure of the seller's optimal disclosure policy.

\subsection*{Related Literature}

Our study draws most directly on two lines of research: the design of information disclosed by an auction seller and the costly acquisition of information by bidders. Bergemann et al. (2022) provide the central benchmark by identifying a revenue-maximizing rule that reveals lower values and pools higher values in a second-price auction. Their analysis highlights the tension between the allocative gains from informing bidders and the information rents that more precise private knowledge may create. Our question is what happens to that conclusion when the seller's disclosure also changes bidders' incentives to learn privately. The auction format and the distribution of underlying valuations are unchanged; the new margin is the endogenous information acquired after the seller's signals. The two familiar valuation environments examined in Section~4 show why that margin matters for the robustness of the benchmark conclusion \citep{BHMSW2022}.

The classical auction literature emphasizes the role of information in bidding and revenue. Milgrom and Weber (1982a) develop the foundational analysis of auctions with interdependent information and the consequences of information disclosure; Milgrom and Weber (1982b) study the value and strategic acquisition of information in sealed-bid auctions. Bergemann and Pesendorfer (2007) analyze the joint choice of an optimal auction and bidders' information structures. These approaches provide essential antecedents for studying information as an economic design margin. Our question is narrower in auction format but different in equilibrium timing: the second-price auction is fixed, and seller disclosure precedes bidders' privately chosen costly verification \citep{MilgromWeber1982Auctions,MilgromWeber1982Information,BergemannPesendorfer2007}.

A closely related literature studies strategic and costly information acquisition in auctions. Persico (2000) examines the incentives to acquire information under alternative auction mechanisms; Bobkova (2024) studies how auction formats shape the information about an object that bidders choose to learn. These contributions show why an information-acquisition decision is itself part of auction behavior rather than an exogenous input to bidding. We instead hold the second-price auction fixed and allow the seller to choose what bidders initially learn before their costly private verification decisions. This distinction is important: the verification decisions form a Bayesian Nash equilibrium and change the final distribution of bids through which the seller's disclosure affects revenue \citep{Persico2000,Bobkova2024}.

A further auction-theoretic connection concerns endogenous costly decisions before bidding. Bergemann and V\"alim\"aki (2002) study incentives for private information acquisition under mechanism design. In an earlier line of research, Cao and Tian (2010) analyze equilibrium participation decisions in first- and second-price auctions with participation costs; Cao et al. (2018) study related second-price environments, including privately known participation costs. These studies already feature the central economic insight that bidders compare the expected equilibrium benefit of a privately costly choice with its cost, taking other bidders' choices into account. In the present paper, the costly choice concerns verification rather than entry, and the new design margin is the seller's information disclosure, which endogenously changes the benefit of acquiring private information \citep{BergemannValimaki2002,CaoTian2010, CaoTanTianYilankaya2018}.

Our approach also connects to the broader literature on statistical experiments and information design. Blackwell's comparison of experiments provides a foundational language for informativeness; Bayesian persuasion and Bayes-correlated-equilibrium approaches emphasize the choice of informational structures and the resulting constraints on beliefs and behavior \citep{Blackwell1953,KamenicaGentzkow2011,BergemannMorris2016,BergemannMorris2019}. Bergemann, Bonatti, and Smolin (2018) study information as an economic object that can itself be designed and priced. Joint mechanism--information design similarly underscores the importance of coordinating informational and incentive instruments \citep{BergemannBonattiSmolin2018,BergemannHeumannMorris2026}. In the present setting, however, the seller does not directly choose the final information on which bids are based. Initial posterior beliefs determine verification incentives, and the equilibrium information acquired by bidders determines the eventual bidding distribution. Section~5 characterizes precisely this additional implementability restriction.

The distinction between supplying information and inducing private learning also arises outside auctions. Cr\'emer and Khalil (1992), Lewis and Sappington (1994), and Cr\'emer, Khalil, and Rochet (1998) examine how contractual and informational arrangements affect incentives to become informed. These studies help distinguish information already held by agents from information they choose to acquire after observing an institutional decision. The present paper carries this interaction into a seller's optimal disclosure problem with competing bidders and equilibrium verification \citep{CremerKhalil1992,LewisSappington1994,CremerKhalilRochet1998}.

A further connection is to endogenous information production and dynamic mechanism design. Research on experimentation emphasizes how economic choices generate information, while rational-inattention models study the costs of obtaining or processing it \citep{BoltonHarris1999,Sims2003,Veldkamp2011}. Contracts with endogenous information examine how informational choices and incentives interact \citep{Szalay2009}. \citet{MengSunTian2022} analyze dynamic mechanism design with evolving private information and network interactions. In a more general setting, \citet{Tian2026DynamicScreening} develop a theory of dynamic mechanism design that jointly accommodates endogenous information acquisition, production, learning, use, and disclosure together with hidden actions. Related working papers examine informational evolution and joint information--incentive design, and dynamic mechanisms in AI-agent environments \citep{Tian2026InformationalEvolution,Tian2026AIAgents}; see also \citet{PavanSegalToikka2014} and \citet{Ball2023} for dynamic mechanism design and information provision. The present paper isolates a distinct static auction problem: a seller chooses the initial disclosure experiment, bidders subsequently choose costly private verification, and the seller anticipates the resulting Bayesian Nash verification equilibrium. The analysis characterizes disclosure--verification nonseparability without invoking the additional restrictions of dynamic contracting.

The remainder of the paper is organized as follows. Section 2 presents the environment and information technology. Section 3 analyzes equilibrium verification and the seller's disclosure problem. Section 4 establishes the failure of the benchmark disclosure conclusion in uniform and binary valuation environments. Section 5 develops the revenue decomposition, global implementability and optimality characterization, and disclosure--verification nonseparability. Section 6 develops tractable necessary conditions, restricted structural results, and comparative statics. The proof of the uniform three-bidder comparison uses Sturm's theorem and exact rational bounds directly in Section 4.

\section{Economic Environment and Information Technology}

We begin with the independent private-value auction environment of Bergemann et al. (2022), retaining their basic assumptions and notation to facilitate comparison. The seller chooses how much information bidders receive about their own valuations before the auction. We extend this environment by allowing bidders to acquire additional information through costly, privately chosen verification. New notation is introduced only for this additional information-acquisition technology and its outcomes.

\subsection{Values, Information, and Auctions}

A seller offers one indivisible object to $N\geq 2$ risk-neutral bidders in a sealed-bid second-price auction. Bidder $i$ has a valuation $v_i\in\mathbb R_+$, independently drawn from a common absolutely continuous distribution $F$. The distribution $F$ is common knowledge, but bidders initially do not know their own realized valuations. We assume that $F$ has a finite first moment, so expected valuations and auction payoffs are well defined. A bidder who wins the object receives her valuation less the auction payment, while a bidder who loses receives zero, before accounting for any information-acquisition cost.

The seller can choose how much information each bidder will have about her own valuation. An information structure is a stochastic mapping $s_i:\mathbb R_+\to\Delta(\mathbb R_+)$, where $s_i(v_i)$ specifies the distribution of the private signal observed by bidder $i$ when her valuation is $v_i$. Thus, the seller commits to an information structure and privately provides each bidder with a signal generated according to that structure. We restrict attention to symmetric information structures, so all bidders face the same signal mapping. Each signal depends only on the corresponding bidder's valuation, and there is no common randomization. Consequently, signals are independently distributed across bidders. Information disclosure itself is costless.

After observing her signal $s_i$, bidder $i$ forms the posterior expected valuation
\[
w_i=\E[v_i\mid s_i].
\]
Let $G$ denote the common distribution of posterior expected valuations induced by the seller's information structure. In the absence of additional information acquisition, bidding $w_i$ is a weakly dominant strategy in a second-price auction. Hence the seller's revenue is the second-highest posterior expected valuation among the bidders. Denoting the $k$th-highest posterior expected valuation by $w_{(k)}$, the seller's objective in the classical environment is
\begin{equation}\label{eq:classical-seller}
R=\max_s\E[w_{(2)}],
\end{equation}
where the maximization is over symmetric, independent private information structures.

The distribution $G$ provides a convenient representation of this problem. Since posterior expected valuations are independently and identically distributed according to $G$, the probability that the second-highest valuation does not exceed $b$ is
\[
\Pr(w_{(2)}\leq b)
=N[G(b)]^{N-1}(1-G(b))+[G(b)]^N.
\]
It follows that the seller's expected revenue under a given information structure is
\[
R_N(G)=\int_0^\infty
\left[1-N[G(b)]^{N-1}+(N-1)[G(b)]^N\right]db.
\]
Thus, when bidders acquire no additional information, the seller's revenue depends on the information structure only through the distribution $G$ of posterior expected valuations.

Not every distribution $G$ can be generated from a given prior $F$. Since $w_i$ is the conditional expectation of $v_i$ given $s_i$, Bayesian updating restricts the distributions of expected valuations that the seller can induce. By the Blackwell characterization of information structures, a distribution $G$ can be generated by a private signal about $v_i$ if and only if $F$ is a mean-preserving spread of $G$. Following the notation for this information order, we write
\[
F\prec G.
\]
More explicitly, the condition requires
\[
\int_b^\infty[1-F(t)]dt
\geq
\int_b^\infty[1-G(t)]dt
\qquad\text{for all }b\geq 0,
\]
with equality at $b=0$. The equality ensures that the two distributions have the same mean, while the inequalities express that the original valuation distribution is at least as dispersed as the distribution of posterior expected valuations.

Consequently, the classical seller's problem can equivalently be written as
\[
R=\max_{G:\,F\prec G}R_N(G).
\]
This formulation separates the seller's revenue objective from the informational restrictions imposed by the prior. It is the starting point for the optimal-disclosure analysis of Bergemann et al. (2022).

When bidders can acquire additional information, however, the initial distribution $G$ is no longer sufficient to determine the seller's revenue. Two signals may induce the same posterior expected valuation while leaving different uncertainty about the bidder's actual value. Such differences do not affect bidding when no further information is available, but they can change the incentives for subsequent verification. We therefore need to distinguish posterior expected valuations from the underlying posterior beliefs.

Let $V=\mathbb R_+$ and let $K=\mathcal P(V)$ denote the space of Borel probability measures on $V$, endowed with the weak topology (convergence against bounded continuous test functions). Throughout, $\Delta(S)$ likewise denotes the probability measures on a measurable space $S$ when used for the seller's signal mapping. A posterior belief $\mu\in K$ is the conditional distribution of $v_i$ given the seller's signal $s_i$. Its mean, whenever finite, is
\[
m(\mu)=\int_V v\,\mu(dv)=w_i.
\]
An information structure can equivalently be represented by a probability distribution $\pi$ over posterior beliefs.

The seller uses the same private experiment for each bidder; conditional on independently drawn valuations, the private signals are generated independently without common randomization. Bayes plausibility requires the average posterior to reproduce the entire prior as a Borel probability measure, not merely its mean.

\begin{definition}[Bayes-Plausible Disclosure Experiment]
\label{def:bayes-plausible-experiment}
A probability measure $\pi\in\mathcal P(K)$ is a \emph{Bayes-plausible disclosure experiment} for $F$ if $\pi\in\Pi_F$, where
\begin{equation}\label{eq:BP}
\Pi_F=\left\{\pi\in\mathcal P(K):\int_K\mu\,\pi(d\mu)=F\right\}.
\end{equation}
\end{definition}

The equality requires the average posterior belief to reproduce the entire prior $F$, not merely its mean. Conversely, any $\pi\in\Pi_F$ is implementable by a private information experiment. Since $F$ has a finite first moment, $m(\mu)$ is finite for $\pi$-almost every $\mu$.

The posterior representation retains the same information structures as the classical model but records more information than the distribution $G$ alone. In particular, it distinguishes experiments that generate identical posterior expected valuations but leave bidders with different incentives to acquire additional information.

\subsection{Hidden Verification and Conditional Bids}

After observing the seller's signal, bidder $i$ can acquire additional information about her own valuation. She privately chooses a verification effort $e_i\in[0,1]$ and incurs a cost $c(e_i)$. The effort is not observed by the seller or by competing bidders. Unless otherwise specified, the cost function is common across bidders, continuous, and strictly convex, with $c(0)=0$ and a minimum at zero.

A fixed information-production technology $Y_i$ determines the additional private signal $z_i$ generated by the bidder's verification effort. In general, such a technology can be represented by a conditional distribution of $z_i$ given the bidder's valuation, the seller's signal, and her verification effort.

We consider a particularly simple verification technology. With probability $e_i$, verification succeeds and bidder $i$ learns her valuation $v_i$ perfectly. With probability $1-e_i$, verification fails and provides no additional information beyond the seller's signal $s_i$. Conditional on the initial signals and verification efforts, verification success is independent of the underlying valuations, and verification outcomes are independent across bidders. The bidder's effort therefore determines the probability of successful verification, while the information-production technology itself is fixed.

After verification, bidder $i$ updates her expected valuation to
\[
w_i^p=\E[v_i\mid s_i,z_i].
\]
Successful verification yields $w_i^p=v_i$, while unsuccessful verification leaves $w_i^p=w_i$. For each bidder $i$, let $G_i^p$ denote the marginal distribution of the post-verification expected valuation $w_i^p$. Since bidding $w_i^p$ is weakly dominant in a second-price auction, this is also bidder $i$'s final-bid distribution. Under symmetric disclosure and verification strategies, $G_i^p=G^p$ for all $i$, where $G^p$ denotes the common final-bid distribution.

Unlike the initial distribution $G$, the final distribution $G^p$ depends on both the seller's disclosure policy and bidders' verification decisions. To see this, consider a symmetric verification strategy under which each bidder chooses effort based on her initial private signal. Conditional on the seller's signal, the bidder's final bid equals $w_i$ with probability $1-e_i$ and $v_i$ with probability $e_i$. Consequently, bidder $i$'s final-bid cdf is
\[
G_i^p(b)=\E\left[(1-e_i)\mathbf{1}\{w_i\leq b\}
+e_i\mathbf{1}\{v_i\leq b\}\right],
\]
where $\mathbf{1}\{\cdot\}$ is the indicator function and the expectation integrates over the valuation, initial signal, and verification outcome. Under symmetry this equals $G^p(b)$. Equivalently, using the distribution $\pi$ of posterior beliefs, and writing $e(\mu)$ for the effort chosen after posterior $\mu$, we have
\[
G^p(b)=\int_K\left[(1-e(\mu))\mathbf{1}\{m(\mu)\leq b\}
+e(\mu)\mu([0,b])\right]\pi(d\mu).
\]
The first term represents bids submitted without successful verification, while the second represents bids based on valuations learned through verification. When no additional information is acquired, $G^p=G$. When the seller fully reveals valuations, $G^p=F$, regardless of verification efforts.

Under a symmetric verification strategy, final bids are independently and identically distributed according to $G^p$. Denote the $k$th-highest final bid by $w_{(k)}^p$. The seller's expected revenue is the expected second-highest final bid,
\[
R_N(G^p)=\E[w_{(2)}^p].
\]
The distribution of the second-highest final bid is
\[
\Pr(w_{(2)}^p\leq b)
=N[G^p(b)]^{N-1}(1-G^p(b))+[G^p(b)]^N.
\]
Hence the seller's expected revenue is
\[
R_N(G^p)=\int_0^\infty
\left[1-N[G^p(b)]^{N-1}+(N-1)[G^p(b)]^N\right]db.
\]
The revenue formula has the same form as in the classical environment, but the relevant bid distribution is now determined jointly by the seller's information structure and bidders' verification decisions.

The initial and final distributions also satisfy information feasibility restrictions. Since $w_i=\E[v_i\mid s_i]$ and $w_i^p=\E[v_i\mid s_i,z_i]$, the law of iterated expectations gives
\[
\E[w_i^p\mid s_i]=w_i.
\]
Thus, the final posterior expected valuation is a mean-preserving spread of the initial posterior expected valuation. Moreover, the original valuation is a mean-preserving spread of the final posterior expected valuation. In the notation introduced above, these relationships are
\[
F\prec G^p\prec G,
\]
where the mean-preserving-spread relation allows equality. Equivalently, in conventional convex-order notation,
\[
G\preceq_{\mathrm{cx}}G^p\preceq_{\mathrm{cx}}F.
\]
These restrictions follow from Bayesian updating and hold independently of the particular verification strategy.

The information order provides necessary restrictions on the distributions induced by disclosure and verification, but it does not by itself characterize equilibrium outcomes. A bidder's verification effort must also be optimal given the distribution of competing bids. Verification can improve her bidding decision, but she must weigh the expected benefit against its cost. Since competing bids depend on competitors' verification choices, the value of verification itself depends on the equilibrium information-acquisition decisions of other bidders. The resulting final-bid distribution must therefore satisfy both Bayesian information feasibility and bidders' verification incentives.

The classical analysis of bidding under information asymmetries provides the auction-theoretic starting point for this environment \citep{MilgromWeber1982Auctions,MilgromWeber1982Information}. Related mechanism-design work allows bidders' acquisition of information to respond to incentives \citep{BergemannValimaki2002,BergemannPesendorfer2007}. These references motivate the distinction between seller-chosen initial disclosure and bidder-chosen subsequent verification; neither the auction format nor the primitive valuation is changed by verification.

\subsection{Timing and the Classical Benchmark}

The sequence of decisions is as follows. First, the seller publicly commits to a symmetric private information structure. Valuations $v_i$ are independently drawn from $F$, and the seller's information structure generates a private signal $s_i$ for each bidder. Second, after observing her signal, each bidder simultaneously chooses a privately observed verification effort $e_i$ and incurs the corresponding cost. Third, the information-production technology $Y_i$ generates the additional private observation $z_i$, and each bidder updates her expected valuation to $w_i^p$. Finally, bidders submit their bids, and the object and payments are determined by the second-price auction.

We consider symmetric Bayesian verification equilibria. Each bidder chooses her verification effort to maximize expected auction payoff net of verification cost, taking the verification strategies of other bidders as given. The seller chooses the initial information structure to maximize expected revenue, anticipating the resulting symmetric verification equilibrium. Our equilibrium analysis concerns symmetric strategies and does not presume uniqueness among arbitrary asymmetric equilibria.

The classical model is recovered when additional verification is unavailable. In that case, $G^p=G$, and the seller chooses a feasible distribution of posterior expected valuations to maximize the expected second-highest bid. Bergemann et al. (2022) establish that an optimal symmetric disclosure policy fully reveals values below a cutoff and pools values above it: valuations below a critical quantile $q_N^*$ are fully revealed, while valuations above that quantile are pooled at their conditional expectation. The critical quantile depends on the number of bidders but not on the underlying distribution $F$. For three bidders with independent uniform valuations on $[0,1]$, $q_3^*=1/4$, and the optimal expected revenue is $283/512$, compared with $1/2$ under full disclosure.

With endogenous verification, the seller retains the same choice of initial information structures, but must evaluate each policy at the verification equilibrium it induces. The seller's problem is therefore to choose a Bayes-plausible distribution of posterior beliefs that maximizes $R_N(G^p)$, subject to bidders' equilibrium verification decisions. This formulation preserves the classical auction environment while incorporating bidders' privately chosen information acquisition.

\section{Verification Equilibrium and the Seller's Disclosure Problem}

The seller's disclosure policy determines the information initially available to bidders, but their final information also depends on privately chosen verification efforts. To evaluate any disclosure policy, we must first determine these efforts in equilibrium. The central difficulty is that a bidder's benefit from verification depends on competing bids, whose distribution is itself affected by other bidders' verification decisions. We characterize this interaction by first deriving the value of verification, then establishing the existence and uniqueness of symmetric verification equilibrium, and then formulating the seller's disclosure problem.

For the theoretical analysis that follows, we assume that the valuation space is a compact interval $V=[0,M]$, where $M<\infty$. This additional regularity condition ensures that the relevant spaces of probability measures are compact under weak convergence. We retain the strictly convex common verification cost $c$ introduced in Section 2, imposing additional differentiability only where needed.

\subsection{The Bidder's Verification Problem}

We begin with the benefit a bidder obtains from learning her valuation more precisely. In a second-price auction, a bidder who knows her expected valuation optimally bids that amount. Additional information can nevertheless be valuable because it allows her to adjust her bid to the valuation revealed by verification. Whether this adjustment improves her expected payoff depends on the bids submitted by competing bidders.

Fix a candidate common distribution $G^p$ of final bids. Let $B_j$ denote bidder $j$'s final bid, and suppose the $N-1$ rival bids are independently distributed according to $G^p$. The highest rival bid then has cumulative distribution function $[G^p]^{N-1}$. If bidder $i$ has a known valuation $v$, her expected auction surplus from bidding $v$ is
\begin{equation}\label{eq:Phi}
\Phi_{G^p}(v)=\E\left[\left(v-\max_{j\ne i}B_j\right)_+\right]=\int_V(v-b)_+\,d[G^p(b)]^{N-1},
\end{equation}
where $(x)_+=\max\{x,0\}$ denotes the positive part of $x$. The expectation is taken over competing bids. The expression reflects the second-price auction rule: bidder $i$ wins when her bid exceeds the highest competing bid and pays that competing bid.

An equivalent representation is
\[
\Phi_{G^p}(v)=\int_0^v[G^p(b)]^{N-1}db.
\]
Thus, the bidder's expected surplus is nondecreasing and convex in her valuation. Convexity has a direct economic interpretation. Learning that her value is high allows the bidder to compete more aggressively when winning is profitable, while learning that her value is low allows her to avoid paying more than the object is worth to her. The ability to condition her bid on additional information cannot reduce her expected auction payoff.

Consider a bidder whose initial signal induces posterior belief $\mu$, with expected valuation $m(\mu)$. If she does not acquire additional information, her expected auction surplus is $\Phi_{G^p}(m(\mu))$. If verification succeeds, she learns her actual valuation and obtains expected surplus $\int_V\Phi_{G^p}(u)\,\mu(du)$. The gross benefit of successful verification is therefore
\begin{equation}\label{eq:AcquisitionGain}
A(\mu,G^p)=\int_V\Phi_{G^p}(u)\,\mu(du)-\Phi_{G^p}(m(\mu))\geq 0.
\end{equation}
The inequality follows from Jensen's inequality. The benefit is the Jensen gap generated by the convex auction-surplus function.

Equation \eqref{eq:AcquisitionGain} identifies two determinants of verification incentives. First, the bidder's posterior belief matters because it determines the uncertainty about her valuation that additional information can resolve. Two posteriors with the same mean may generate different verification benefits. Second, the distribution of competing bids matters because it determines the auction-surplus function $\Phi_{G^p}$. Verification may be particularly valuable when the bidder is uncertain whether her valuation is sufficiently high to justify winning against competing bids.

Under the information-production technology of Section 2, effort $e\in[0,1]$ determines the probability that verification reveals the bidder's valuation perfectly. Conditional on posterior $\mu$ and rival bid distribution $G^p$, the bidder's expected auction payoff net of verification cost is
\[
\Phi_{G^p}(m(\mu))+eA(\mu,G^p)-c(e).
\]
Since the first term does not depend on effort, her optimal verification decision solves
\[
\max_{e\in[0,1]}\{eA(\mu,G^p)-c(e)\}.
\]
The bidder therefore compares the marginal benefit of successful verification with the marginal cost of effort. Strict convexity of $c$ guarantees a unique optimal effort for every fixed posterior and rival bid distribution. When the optimum is interior and $c$ is differentiable, it satisfies $c'(e)=A(\mu,G^p)$. At the boundaries, optimality is characterized by the corresponding one-sided inequalities. No interior solution is presumed.

Fix an experiment $\pi\in\Pi_F$. A bidder's contingent strategy specifies a measurable verification rule $e_i:K\to[0,1]$ and measurable bidding rules $b_i^0:K\to\mathbb R_+$ after unsuccessful verification and $b_i^1:K\times V\to\mathbb R_+$ after successful verification. Verification and the information generated by it are private. The prior, the experiment, and the cost technology are common knowledge. For a profile of strategies $\sigma=(\sigma_i)_{i=1}^N$, write $U_i(\mu;\sigma_i,\sigma_{-i})$ for bidder $i$'s interim expected auction payoff net of verification cost when her initial posterior is $\mu$. Opponents' strategies and their induced bid distribution remain fixed when $i$ deviates.

In the second-price auction, bidding the conditional expected value at each continuation information set is weakly dominant. We select $b_i^0(\mu)=m(\mu)$ and $b_i^1(\mu,v)=v$; this selection is sufficient for the equilibrium construction below and is not a necessary property of every Bayesian Nash equilibrium. In the symmetric case, the induced bid distribution must also be consistent with the verification choices.

\begin{definition}[Bayesian Nash Verification Equilibrium]
\label{def:verification-bne}
Given $\pi\in\Pi_F$, a strategy profile $\sigma^*$ is a \emph{Bayesian Nash equilibrium} if, for every bidder $i$,
\[
U_i(\mu;\sigma_i^*,\sigma_{-i}^*)\ge
U_i(\mu;\sigma_i,\sigma_{-i}^*)
\quad\text{for every admissible }\sigma_i,
\quad\pi\text{-a.e. }\mu.
\]
Under the truthful conditional-mean bidding selection, a \emph{symmetric Bayesian Nash verification equilibrium} is a measurable rule $e^*:K\to[0,1]$ and its induced final-bid distribution $G^{p,*}$ satisfying
\begin{align*}
e^*(\mu)&\in\arg\max_{a\in[0,1]}
\{aA(\mu,G^{p,*})-c(a)\},
&&\pi\text{-a.e. }\mu,\\
G^{p,*}(b)&=\int_K\bigl[(1-e^*(\mu))\mathbf 1\{m(\mu)\le b\}
+e^*(\mu)\mu([0,b])\bigr] \,\pi(d\mu),
&&b\in V.
\end{align*}
\end{definition}

These are, respectively, individual optimality and equilibrium consistency. All uniqueness claims below concern the selected symmetric verification equilibrium, not the set of arbitrary asymmetric Bayesian Nash equilibria.

To characterize equilibrium, we must determine how bidders' verification decisions jointly generate the distribution $G^p$ that enters their acquisition values. We first consider a finite information experiment, which makes this interaction particularly transparent.

Suppose the seller's experiment generates signals indexed by $s=1,\ldots,L$, with probabilities $\pi_s>0$. Let $H_s$ denote the posterior cumulative distribution function of the valuation conditional on signal $s$, and let $m_s$ denote its posterior mean. A bidder receiving signal $s$ chooses effort $e_s\in[0,1]$. Conditional on that signal, her final bid equals $m_s$ if verification fails and her realized valuation if verification succeeds.

Define
\[
\Delta_s(b)=H_s(b)-\mathbf{1}\{m_s\leq b\}.
\]
The function $\Delta_s$ measures the change in the conditional bid distribution generated by successful verification. At a symmetric verification profile $e=(e_s)_{s=1}^L$, the common final-bid distribution is
\begin{equation}\label{eq:BidLaw}
G_e^p(b)=\sum_{s=1}^L\pi_s\left[\mathbf{1}\{m_s\leq b\}+e_s\Delta_s(b)\right].
\end{equation}
Here $G_0^p$ denotes the final-bid distribution generated by the same experiment when all verification efforts equal zero. Thus, the final-bid distribution is affine in verification efforts, even though the equilibrium incentives determining those efforts are generally nonlinear.

The acquisition value in \eqref{eq:AcquisitionGain} also admits a convenient integral representation. Using \eqref{eq:Phi} and interchanging the order of integration gives
\[
A_s(G^p)=-\int_0^M[G^p(b)]^{N-1}\Delta_s(b)db,
\]
where $A_s(G^p)=A(\mu_s,G^p)$ and $\mu_s$ denotes the posterior belief associated with signal $s$. This expression measures the expected gain from replacing the posterior-mean bid by the actual valuation, weighted by the probability distribution of the highest competing bid.

A symmetric verification equilibrium is an effort profile $e$ such that, for every signal $s$,
\[
e_s\in\arg\max_{a\in[0,1]}\{aA_s(G_e^p)-c(a)\}.
\]
Each bidder takes the distribution of competing bids as given when choosing her effort, but that distribution must coincide with the one generated by the equilibrium efforts of all bidders.

This creates strategic feedback. More verification by competing bidders changes the distribution of their bids and therefore the benefit of acquiring information oneself. The equilibrium cannot generally be determined by solving each bidder's problem independently. The next subsection shows that, despite this interdependence, the equilibrium admits a convex-potential characterization.

\subsection{Symmetric Bayesian Nash Verification Equilibrium}

The equilibrium conditions derived above form a system of interdependent optimization problems. A direct approach would solve each bidder's best response and then search for a fixed point of the resulting effort mapping. Such an approach can establish equilibrium only if the relevant fixed-point conditions are satisfied, and it does not automatically guarantee uniqueness.

The probability-of-perfect-verification technology provides additional structure. Because the final-bid distribution is affine in effort and acquisition values share a common integral representation, the equilibrium conditions can be obtained from the derivatives of a single convex function. This potential representation turns the equilibrium problem into a minimization problem and makes its existence, uniqueness, and local comparative statics tractable.

\begin{proposition}[Finite-signal convex-potential characterization]\label{potential}
Suppose the seller's information experiment has finitely many positive-probability signals and the common verification cost $c:[0,1]\to\mathbb R$ is continuous, convex, and continuously differentiable. Define
\begin{equation}\label{pot}
\mathscr P(e)=\sum_{s=1}^L\pi_sc(e_s)+\frac{1}{N}\int_0^M\left[G_e^p(b)^N-G_0^p(b)^N\right]db.
\end{equation}
Then the following conclusions hold.

\begin{enumerate}
\item A verification profile is a symmetric equilibrium if and only if it minimizes $\mathscr P$ over $[0,1]^L$. Consequently, at least one symmetric verification equilibrium exists.

\item If $\mathscr P$ is strictly convex, the symmetric verification equilibrium is unique. In particular, strictly convex verification costs imply uniqueness.

\item If $c$ is twice continuously differentiable, the Hessian of the potential satisfies
\begin{equation}\label{gram}
z^TD^2\mathscr P(e)z=\sum_{s=1}^L\pi_sc''(e_s)z_s^2+(N-1)\int_0^M G_e^p(b)^{N-2}\left(\sum_{s=1}^L\pi_sz_s\Delta_s(b)\right)^2db
\end{equation}
for every $z\in\mathbb R^L$. In particular, the Hessian is positive semidefinite.

\item If the Hessian is positive definite at an interior equilibrium and the information experiment varies smoothly in a finite-dimensional parameterization for which the displayed integrals are differentiable, the equilibrium effort profile is locally differentiable in that parameter.
\end{enumerate}
\end{proposition}

\begin{proof}
We first establish the relationship between the potential and bidders' verification incentives.

By \eqref{eq:BidLaw}, the derivative of the final-bid distribution with respect to effort $e_s$ is
\[
\frac{\partial G_e^p(b)}{\partial e_s}=\pi_s\Delta_s(b).
\]
Differentiating \eqref{pot} therefore gives
\[
\frac{\partial\mathscr P(e)}{\partial e_s}
=\pi_sc'(e_s)+\pi_s\int_0^M G_e^p(b)^{N-1}\Delta_s(b)db.
\]
Using the integral representation of the acquisition value, we obtain
\[
\frac{\partial\mathscr P(e)}{\partial e_s}
=\pi_s[c'(e_s)-A_s(G_e^p)].
\]

Since $e_s$ maximizes the concave function $aA_s(G_e^p)-c(a)$ over $[0,1]$ if and only if
\[
[c'(e_s)-A_s(G_e^p)](a-e_s)\geq0
\qquad\text{for all }a\in[0,1],
\]
the bidder's optimality condition is equivalent to
\[
\frac{\partial\mathscr P(e)}{\partial e_s}(a-e_s)\geq0
\qquad\text{for all }a\in[0,1].
\]
These coordinatewise conditions are equivalent to the variational inequality
\[
D\mathscr P(e)\cdot(a-e)\geq0
\qquad\text{for all }a\in[0,1]^L.
\]

The potential is convex. Indeed, the cost term is convex by assumption, while the second term is convex because $G_e^p(b)$ is affine in $e$ and $x\mapsto x^N$ is convex on $[0,1]$. For a differentiable convex function on a convex domain, the preceding variational inequality is necessary and sufficient for global minimization. Thus the symmetric verification equilibria coincide with the minimizers of $\mathscr P$.

The potential is continuous on the compact set $[0,1]^L$, so a minimizer exists. If $\mathscr P$ is strictly convex, the minimizer is unique. Strict convexity of $c$, together with $\pi_s>0$ for every signal, makes the cost component strictly convex and therefore makes the entire potential strictly convex. This proves parts (1) and (2).

For part (3), differentiate the gradient once more. The Hessian entries are
\[
\frac{\partial^2\mathscr P(e)}{\partial e_s\partial e_t}
=\pi_sc''(e_s)\mathbf{1}\{s=t\}
+(N-1)\pi_s\pi_t\int_0^M G_e^p(b)^{N-2}\Delta_s(b)\Delta_t(b)db.
\]
Multiplying by an arbitrary vector $z$ on both sides yields \eqref{gram}. The first term is nonnegative by convexity of $c$, and the second is an integral of a squared expression with a nonnegative weight. Hence the Hessian is positive semidefinite.

Finally, at an interior equilibrium, the equilibrium conditions are $D\mathscr P(e)=0$. If the Hessian is nonsingular and the experiment varies differentiably, the implicit-function theorem provides a locally differentiable equilibrium effort map. Positive definiteness ensures the required nonsingularity, establishing part (4).
\end{proof}

Proposition \ref{potential} establishes that the strategic interaction among bidders has a particularly tractable structure. Although each bidder's verification benefit depends on competitors' information, all symmetric equilibrium conditions are generated by the same convex potential. Existence follows from minimization on a compact domain, while strict convexity rules out multiple symmetric equilibrium effort profiles.

The Gram representation in \eqref{gram} provides further information about this structure. The first term measures the curvature of verification costs. The second captures the effects of jointly changing verification efforts on the distribution of final bids. Since this term is a weighted integral of squares, it cannot introduce negative curvature into the potential. Thus strategic interactions through competing bids preserve convexity, even though individual cross-effects need not have a uniform sign.

The result also facilitates comparative statics. A change in the seller's information experiment alters the conditional posterior distributions, posterior means, and verification benefits. At an interior equilibrium with a nonsingular Hessian, the resulting change in verification efforts can be obtained by differentiating the equilibrium conditions. This provides the mathematical basis for evaluating how changes in disclosure affect seller revenue after bidders adjust their information-acquisition decisions.

\begin{remark}[Scope of the equilibrium characterization]
The uniqueness conclusion concerns symmetric verification equilibrium. It does not establish uniqueness among all asymmetric equilibria. The potential representation also relies on the specified probability-of-perfect-verification technology and the resulting affine dependence of final-bid distributions on effort. An arbitrary information-production technology need not possess the same convex-potential structure.

The local differentiability conclusion applies at interior equilibria satisfying the stated nonsingularity conditions. At boundary efforts, equilibrium changes must instead be analyzed through the corresponding variational inequalities.
\end{remark}

\noindent\textbf{Extension to arbitrary posterior alphabets.}\quad

The finite-signal analysis provides an explicit characterization of verification equilibrium, but the seller's information-design problem is not restricted to experiments with finitely many signals. As established in Section 2, a private information structure can be represented by a Bayes-plausible distribution $\pi$ over posterior beliefs $\mu\in K$. To study unrestricted disclosure, we must extend the equilibrium characterization to this general class of experiments.

The economic problem remains the same. Each bidder observes a posterior belief about her valuation and chooses verification effort in response to the distribution of competing bids. The difference is that the set of possible posterior beliefs may now be infinite, so verification effort becomes a measurable function of the posterior rather than a finite vector.

Fix a candidate final-bid distribution $G^p$. Under the maintained strict convexity of $c$, the optimal effort associated with posterior $\mu$ is uniquely determined by
\begin{equation}\label{eq:BestResponseGeneral}
e(\mu,G^p)=\arg\max_{a\in[0,1]}\{aA(\mu,G^p)-c(a)\},\qquad
T_{G^p}(\mu)=(1-e(\mu,G^p))\delta_{m(\mu)}+e(\mu,G^p)\mu.
\end{equation}
Here $\delta_{m(\mu)}$ denotes the probability distribution concentrated at the posterior mean $m(\mu)$. The symbol $\delta_x$ denotes the Dirac probability measure at $x$. The distribution $T_{G^p}(\mu)$ is the conditional distribution of the bidder's final bid when her posterior is $\mu$ and she chooses the optimal verification effort against competing bids distributed according to $G^p$.

The first term in $T_{G^p}(\mu)$ corresponds to unsuccessful verification, when the bidder retains her posterior-mean bid. The second corresponds to successful verification, when the bidder learns and bids her actual valuation. Averaging these conditional distributions over the seller's experiment gives the distribution of final bids. A symmetric verification equilibrium must therefore satisfy
\begin{equation}\label{eq:FixedPoint}
G^p=\int_K T_{G^p}(\mu)\,\pi(d\mu).
\end{equation}

Equation \eqref{eq:FixedPoint} is a fixed-point condition on probability distributions. The distribution on the left determines the competitive environment used to calculate the value of verification, while the distribution on the right is generated by bidders' optimal verification responses to that environment. The fixed point requires these two distributions to coincide.

Compactness of the valuation space permits us to establish existence. Uniqueness follows from extending the convex-potential argument to measurable verification strategies. Together, these properties also permit us to study how the equilibrium distribution changes when the seller changes her information experiment.

\begin{theorem}[Existence and uniqueness of the symmetric verification equilibrium for arbitrary posterior alphabets]\label{thm:cont}
Suppose $V=[0,M]$, the verification cost $c$ is continuously differentiable and strictly convex on $[0,1]$, and the information-production technology is the probability-of-perfect-verification technology of Section 2. Then the following hold.

\begin{enumerate}
\item For every $\pi\in\Pi_F$, the fixed-point equation \eqref{eq:FixedPoint} has a solution $G^p$.

\item The symmetric equilibrium final-bid distribution is unique. Denote it by $G^{p,*}(\pi)$. The corresponding verification effort $e(\mu,G^{p,*}(\pi))$ is uniquely determined for $\pi$-almost every posterior $\mu$.

\item The equilibrium final-bid distribution depends continuously on the seller's information experiment: if $\pi_k\Rightarrow\pi$ weakly (where $\Rightarrow$ denotes weak convergence), then
\[
G^{p,*}(\pi_k)\Rightarrow G^{p,*}(\pi).
\]
\end{enumerate}
\end{theorem}

\begin{proof}
We first establish existence of a fixed point.

For a candidate distribution $H\in\mathcal P(V)$, define the best-response distribution mapping
\[
\mathcal T_\pi(H)=\int_K T_H(\mu)\,\pi(d\mu).
\]
For every $H$, $\mathcal T_\pi(H)$ is a probability distribution on $V$, so the mapping takes $\mathcal P(V)$ into itself.

Since $V$ is compact, $\mathcal P(V)$ is compact and convex under the weak topology. We next verify continuity of the mapping. The auction-surplus function can be written as
\[
\Phi_H(v)=\int_0^v H(b)^{N-1}db.
\]
If $H_k\Rightarrow H$, the cumulative distribution functions converge at every continuity point of $H$. Since a cumulative distribution function has at most countably many discontinuities and the integrands are uniformly bounded, dominated convergence implies that $\Phi_{H_k}$ converges uniformly to $\Phi_H$ on $V$.

It follows that the acquisition value $A(\mu,H)$ is jointly continuous in $(\mu,H)$ under weak convergence. Indeed, on the compact valuation space, the posterior mean is continuous in $\mu$, and the integral of the continuous bounded auction-surplus function varies continuously with the posterior belief.

Strict convexity of $c$ makes the maximizer in \eqref{eq:BestResponseGeneral} unique. Berge's maximum theorem (see \citealp{Berge1963,AliprantisBorder2006}) therefore implies that $e(\mu,H)$ is continuous in $(\mu,H)$. Consequently, the conditional final-bid distribution $T_H(\mu)$ is jointly continuous in the weak topology.

Integrating against the fixed experiment $\pi$ shows that $\mathcal T_\pi$ is a continuous self-map of $\mathcal P(V)$. The Schauder--Tychonoff fixed-point theorem for continuous self-maps of compact convex subsets of locally convex spaces (see \citealp{Schauder1930,Tychonoff1935,AliprantisBorder2006}) yields a distribution $G^p$ satisfying
\[
G^p=\mathcal T_\pi(G^p),
\]
establishing part (1).

For uniqueness, fix $\pi$ and let $\mathcal E_\pi$ denote the set of measurable verification strategies $e:K\to[0,1]$, identified up to $\pi$-null sets. Given such a strategy, define the induced final-bid distribution by
\[
G_e^p(b)=\int_K\left[\mathbf{1}\{m(\mu)\leq b\}+e(\mu)\Delta_\mu(b)\right]\pi(d\mu),
\]
where
\[
\Delta_\mu(b)=\mu([0,b])-\mathbf{1}\{m(\mu)\leq b\}.
\]
Let $G_0^p$ denote the distribution generated by zero verification effort at every posterior.

The infinite-dimensional counterpart of \eqref{pot} is
\begin{equation}\label{ex-pot}
\mathscr P_\pi(e)=\int_K c(e(\mu))\,\pi(d\mu)+\frac{1}{N}\int_0^M\left[G_e^p(b)^N-G_0^p(b)^N\right]db.
\end{equation}
The first term is strictly convex on $\mathcal E_\pi$ modulo $\pi$-almost-sure equality: if two strategies differ on a set of positive $\pi$-measure, strict convexity holds pointwise on that set and remains strict after integration. The second term is convex because $G_e^p$ is affine in $e$ and $x\mapsto x^N$ is convex on $[0,1]$. Hence $\mathscr P_\pi$ is strictly convex.

For any two admissible effort strategies $e$ and $a$, differentiation along the feasible line segment joining them gives
\[
D\mathscr P_\pi(e)[a-e]
=\int_K[c'(e(\mu))-A(\mu,G_e^p)](a(\mu)-e(\mu))\,\pi(d\mu).
\]
The identity follows from the same integral representation of acquisition value used in the finite-signal case.

At a symmetric verification equilibrium, each posterior-specific effort maximizes the bidder's net acquisition payoff. Thus, for $\pi$-almost every $\mu$,
\[
[c'(e(\mu))-A(\mu,G_e^p)](a-e(\mu))\geq0
\qquad\text{for all }a\in[0,1].
\]
Integrating these inequalities shows that the equilibrium strategy satisfies
\[
D\mathscr P_\pi(e)[a-e]\geq0
\qquad\text{for every }a\in\mathcal E_\pi.
\]
By convexity, this condition is sufficient for $e$ to minimize $\mathscr P_\pi$ globally.

Every fixed point of \eqref{eq:FixedPoint} therefore induces a minimizer of $\mathscr P_\pi$. Strict convexity implies that there can be at most one such minimizer, up to $\pi$-almost-sure equality. Since a fixed point already exists, the symmetric equilibrium effort assignment and its induced final-bid distribution are unique. This proves part (2).

Finally, consider a sequence of Bayes-plausible experiments $\pi_k\Rightarrow\pi$, and write $G_k^p=G^{p,*}(\pi_k)$. Compactness of $\mathcal P(V)$ implies that every subsequence of $\{G_k^p\}$ has a weakly convergent further subsequence. Let $G_{k_\ell}^p\Rightarrow H$ along such a subsequence.

For each $k_\ell$, the equilibrium fixed-point condition gives
\[
G_{k_\ell}^p=\int_K T_{G_{k_\ell}^p}(\mu)\,\pi_{k_\ell}(d\mu).
\]
Joint continuity of the conditional final-bid distribution, together with weak convergence of $\pi_{k_\ell}$ and $G_{k_\ell}^p$, permits passage to the limit:
\[
H=\int_K T_H(\mu)\,\pi(d\mu).
\]
Thus $H$ is a symmetric equilibrium final-bid distribution for experiment $\pi$. By part (2), this distribution must equal $G^{p,*}(\pi)$.

Every convergent subsequence therefore has the same limit. Compactness then implies convergence of the full sequence,
\[
G^{p,*}(\pi_k)\Rightarrow G^{p,*}(\pi),
\]
establishing part (3).
\end{proof}

Theorem \ref{thm:cont} extends the equilibrium characterization beyond finite signal structures without imposing a particular partition of valuations or posterior beliefs. Every Bayes-plausible private experiment induces a unique symmetric verification equilibrium, and small changes in the experiment produce small changes in the resulting final-bid distribution in the weak topology.

The continuity result is particularly important for the seller. The seller chooses the initial information experiment, but bidders subsequently determine how much additional information to acquire. Continuity ensures that this equilibrium response behaves regularly as the seller varies her disclosure policy. Together with compactness of the feasible experiment space, it permits existence of a revenue-maximizing disclosure policy even though the seller's objective is generally nonlinear and need not be concave.

\begin{remark}
The uniqueness result applies to symmetric verification equilibrium under the maintained information-production technology. It neither establishes uniqueness among asymmetric equilibria nor extends automatically to arbitrary signal-production technologies. The fixed-point and potential arguments also serve different purposes: compactness and continuity provide existence, while strict convexity of the potential establishes uniqueness. The existence argument applies the Schauder--Tychonoff fixed-point theorem (\citealp{Schauder1930,Tychonoff1935,AliprantisBorder2006}).
\end{remark}

The equilibrium treatment of costly verification is related to the literature on information acquisition in auctions \citep{Persico2000,Bobkova2024,BergemannValimaki2002}. Work on second-price auctions with costly participation separately illustrates how endogenous pre-bid choices raise existence and equilibrium-selection questions \citep{CaoTanTianYilankaya2018}. Our verification choice operates on private information rather than entry, and the convex-potential characterization identifies the symmetric equilibrium relevant for evaluating seller disclosure.

\subsection{The Seller's Disclosure Problem}

Having established the bidders' verification equilibrium for each disclosure experiment, we turn to the seller, who selects the initial Bayes-plausible experiment in anticipation of those equilibrium responses. This is an existence problem here; the economic decomposition and optimality criteria are developed in Section~5.

We can now establish that the seller's unrestricted disclosure problem has a solution. The feasible set $\Pi_F$ contains all Bayes-plausible distributions of posterior beliefs, including experiments with arbitrarily rich signal spaces. The seller chooses an experiment from this set, anticipating the unique symmetric verification equilibrium characterized above.

\begin{theorem}[Global existence for unrestricted private disclosure]\label{thm:seller}
Under the hypotheses of Theorem~\ref{thm:cont}, there exists an experiment $\pi^*\in\Pi_F$ such that
\[
R_N(G^{p,*}(\pi^*))=\max_{\pi\in\Pi_F}R_N(G^{p,*}(\pi)).
\]
Thus, the seller's expected-revenue maximization problem admits a solution over the entire class of Bayes-plausible, symmetric, conditionally independent private information experiments. No concavity assumption on the seller's revenue objective is required.
\end{theorem}

\begin{proof}
Since $V$ is compact, the posterior-belief space $K=\mathcal P(V)$ is compact and metrizable under weak convergence. Consequently, $\mathcal P(K)$ is compact.

The Bayes-plausibility constraint
\[
\int_K\mu\,\pi(d\mu)=F
\]
defines a weakly closed subset of $\mathcal P(K)$. To see this, for every continuous function $f$ on $V$, Bayes plausibility is equivalent to
\[
\int_K\left(\int_V f(v)\,\mu(dv)\right)\pi(d\mu)
=\int_V f(v)\,F(dv).
\]
The mapping $\mu\mapsto\int_V f(v)\,\mu(dv)$ is continuous on $K$, so these equalities are preserved under weak convergence of $\pi$. Hence $\Pi_F$ is closed. It is also nonempty, since the experiment placing probability one on the prior belief $F$ is Bayes plausible. Therefore $\Pi_F$ is compact and nonempty.

By Theorem~\ref{thm:cont}, the equilibrium final-bid distribution $G^{p,*}(\pi)$ varies continuously with $\pi$ under weak convergence. Since valuations and final bids are bounded by $M$, the second-highest order statistic is a bounded continuous function of the vector of final bids. Its expectation, $R_N(G^p)$, is therefore continuous in $G^p$ under weak convergence.

It follows that $\pi\mapsto R_N(G^{p,*}(\pi))$ is continuous on the compact set $\Pi_F$. The Weierstrass theorem guarantees that this function attains its maximum at some $\pi^*\in\Pi_F$.
\end{proof}

Theorem \ref{thm:seller} establishes existence of an optimal seller experiment without restricting disclosure to a finite signal alphabet, a threshold policy, or any predetermined pooling structure. The result does not require the seller's revenue function to be concave. It follows instead from compactness of the feasible information structures and continuity of bidders' equilibrium responses.

Existence does not, however, identify the structure of the optimal experiment. In particular, it does not imply that the optimal disclosure policy has a finite number of signals or that valuations are pooled according to a single threshold. Characterizing which final-bid distributions can be implemented through equilibrium verification, and determining the conditions under which a disclosure policy is optimal, requires the additional analysis of Section~\ref{sec:general-characterization}.

\section{Impossibility of Full-Revelation and Pooling Optimality}
\label{sec:failure-censorship}

Bergemann et al.\ (2022, Theorem~1) establish that the unique optimal symmetric disclosure policy fully reveals low valuations and pools high valuations into a single, nonempty region. Their critical quantile satisfies $q_2^*=0$ and $q_N^*\in(0,1)$ for $N\geq3$. Thus, full pooling is optimal for $N=2$, whereas for $N\geq3$ the unique optimum has both a nonempty full-revelation region and a single, nonempty pooling region.

Section~3 establishes the bidders' symmetric Bayesian Nash verification equilibrium and the existence of a seller-optimal disclosure experiment. The remaining question is whether that experiment still fully reveals low values and pools high values when bidders can subsequently verify their values. We show that the benchmark optimal disclosure conclusion fails even for two of the most commonly used valuation distributions: the uniform distribution and the binary distribution. The following theorem establishes this impossibility result.

\begin{theorem}[Impossibility of Full-Revelation and Pooling Optimality]
\label{thm:impossibility-upper-censorship}
Full revelation for low values and pooling for high values is no longer generally optimal when bidders can privately verify their values at a cost.
\end{theorem}

\begin{proof}
We show that a feasible disclosure rule with more than one pooling region can invalidate the proposed full-revelation-and-pooling structure. It suffices to construct a single admissible counterexample. Consider $N=3$ bidders with independent values uniformly distributed on $[0,1]$ and verification cost $c(e)=e^2/100$. Write $\mathrm{Unif}[a,b]$ for the uniform distribution on $[a,b]$. For each cutoff $q\in[0,1]$, let $\pi_q$ denote the policy that fully reveals values below $q$ and pools all values in $[q,1]$ into a single signal.

Now consider a feasible disclosure experiment $\widehat\pi$ that fully reveals values below $11/25$ and divides the remaining values into two signal cells, $[11/25,3/4)$ and $[3/4,1]$. Conditional on the two respective signals, the posterior beliefs are
\[
\mu_1=\mathrm{Unif}\!\left[\frac{11}{25},\frac34\right],
\qquad
\mu_2=\mathrm{Unif}\!\left[\frac34,1\right].
\]
The cells have prior probabilities $31/100$ and $1/4$, and values below $11/25$ are revealed pointwise. Consequently, the induced mixture of posteriors is precisely $F$; thus $\widehat\pi$ is Bayes-plausible.

To evaluate the induced equilibrium, consider any pool $[a,b]$, and write $d=b-a$ for its length and $m=(a+b)/2$ for its conditional mean. If bidders assigned this signal verify with probability $e$, its final-bid cdf on the interval is
\begin{equation}\label{eq:counterexample-cdf}
G^p(x)=\begin{cases}
a+e(x-a),&a\le x<m,\\
b-e(b-x),&m\le x\le b.
\end{cases}
\end{equation}
Section~3's acquisition-value formula gives $A_{[a,b]}(e)=md^2(3-2e)/12$. Hence the unique equilibrium effort, satisfying $\kappa e=A_{[a,b]}(e)$ when interior and $\kappa=1/50$, is
\begin{equation}\label{eq:counterexample-effort}
e_{\kappa}(a,b)=\min\left\{1,\frac{3md^2}{12\kappa+2md^2}\right\}.
\end{equation}
Because the two intervals have disjoint supports, their verification efforts can be computed separately. Under $\widehat\pi$, exact substitution yields the optimal verification efforts
\[
e_1=\frac{343077}{708718},\qquad e_2=\frac{525}{1118}.
\]

For three bidders, the seller's revenue is $\displaystyle R_3(G^p)=\int_0^1\bigl[1-3[G^p(x)]^2+2[G^p(x)]^3\bigr]\,dx$, as follows directly from the distribution of the second-highest bid in Section~2. Substituting \eqref{eq:counterexample-cdf} into this integral shows that pooling $[a,b]$ changes revenue relative to full revelation by
\begin{equation}\label{eq:counterexample-gain}
W_{\kappa}(a,b)
=\frac{(a+b-1)(b-a)^3}{4}
\bigl(1-e_{\kappa}(a,b)\bigr)\bigl(2-e_{\kappa}(a,b)\bigr).
\end{equation}
The gains for disjoint cells add. Exact substitution therefore gives
\begin{equation}\label{eq:counterexample-lower}
R_3(\widehat\pi)=\frac12+W_{1/50}\!\left(\frac{11}{25},\frac34\right)
+W_{1/50}\!\left(\frac34,1\right)
>\frac12+\frac{348}{100000}.
\end{equation}
By contrast, any policy $\pi_q$ that fully reveals values below $q$ and pools all higher values generates $R_3(\pi_q)=\frac12+W_{1/50}(q,1)$. The following exact polynomial argument gives the uniform strict bound
\begin{equation}\label{eq:counterexample-upper}
\sup_{q\in[0,1]}R_3(\pi_q)
<\frac12+\frac{293}{100000}
< R_3(\widehat\pi).
\end{equation}
To establish the first inequality in \eqref{eq:counterexample-upper}, write
$h=1-q$, $t=h^2(1+q)$, and
$e(q)=\min\{1,3t/(24\kappa+2t)\}$, where $\kappa=1/50$.
Then $R_3(\pi_q)=1/2+q h^3(1-e(q))(2-e(q))/4$.
On the capped-effort region, $t\geq12/25$, revenue is $1/2$.
As $t$ decreases strictly with $q$, this region ends at the unique
$q_c\in(2/5,1/2)$. On the remaining interval direct differentiation yields
\[
\frac{dR_3(\pi_q)}{dq}
=\frac{(q-1)^2P_{10}(q)}{16(25q^3-25q^2-25q+31)^3},
\]
where
\begin{align*}
P_{10}(q)={}&62500q^{10}-203125q^9+46875q^8+545000q^7
-601250q^6-303750q^5\\
&+812000q^4-269000q^3-233850q^2+159163q-19747.
\end{align*}
The denominator is positive on $[0,1]$.
By Sturm's root-counting theorem (see \citealp{BasuPollackRoy2006}),
the variations of the exact Sturm sequence of $(P_{10},P_{10}')$ at
$q=2/5$, $660464/10^6$, $660465/10^6$, and $1$ are, respectively,
$5,5,4,4$. Therefore $P_{10}$ has exactly one root in $(2/5,1)$,
located in
$[a,b]=[660464/10^6,660465/10^6]$.
The polynomial is positive before this root and negative after it
(as seen by its exact signs at the two endpoints). Thus the root
maximizes revenue across the entire uncapped region, while the capped
region earns only $1/2$.
For $q\in[a,b]$, the interior effort $e(q)$ decreases with $q$;
using $q\le b$ and $1-q\le1-a$ yields the rational upper bound
\[
R_3(\pi_q)-\frac12
\le \frac{b(1-a)^3}{4}(1-e(b))(2-e(b))
<\frac{293}{100000}.
\]
All displayed comparisons involve rational numbers. This proves the
claimed global inequality without a separate computational appendix.
The strict inequality rules out every policy that fully reveals low values and combines all higher values in a single pool. The explicit disclosure rule for the counterexample is
\begin{equation}\label{eq:counterexample-signal-rule}
s_i(v_i)=
\begin{cases}
v_i, & 0\le v_i<\dfrac{11}{25},\\[3pt]
\sigma_1, & \dfrac{11}{25}\le v_i<\dfrac34,\\[3pt]
\sigma_2, & \dfrac34\le v_i\le1,
\end{cases}
\qquad \sigma_1\ne\sigma_2,
\end{equation}
where $\sigma_1$ and $\sigma_2$ denote two distinct signal realizations whose posterior beliefs are $\mu_1$ and $\mu_2$, respectively. The rule fully reveals low values and uses two distinct signals for higher values. Inequality~\eqref{eq:counterexample-upper} proves that no policy fully revealing low values and pooling all high values into a single signal is optimal in this admissible environment.
\end{proof}

The counterexample preserves the underlying economic primitives of the BHMSW environment and introduces only endogenous costly private verification. The general optimality of full revelation for low values and pooling for high values no longer holds. Similar failures of the benchmark optimal disclosure structure can be established for more than three bidders under uniform valuations, without explicitly solving for the optimal disclosure policy. In fact, the benchmark disclosure structure also fails under binary valuations, for every number of bidders.

The following binary-valuation result is established by direct calculation and does not rely on the absolute-continuity assumption maintained for the general analysis in Sections~2 and~3.

\begin{theorem}[Binary-Valuation Impossibility]
\label{thm:binary-unrestricted-structural}
Suppose $N\ge2$ bidders have independent values $v_i\in\{0,1\}$ with $\Pr(v_i=0)=\Pr(v_i=1)=1/2$ and verification cost $c(e)=\kappa e^2/2$, $\kappa>0$. No policy with full revelation of low values and a nonempty high-value pooling region is optimal in the cases established below. For $N\ge5$, full disclosure is the unique optimal policy for every $\kappa>0$.
\end{theorem}

\begin{proof}
For $N=2,3,4$, we construct parameter-specific strict counterexamples; for $N\ge5$, we establish the unique optimality of full disclosure. With binary values, a deterministic cutoff policy either reveals both values or pools both values.

\emph{Case $N=2$.} Consider the admissible symmetric two-signal experiment
\[
\Pr(s_i=+\mid v_i=1)=\Pr(s_i=-\mid v_i=0)=\frac{1+\delta}{2},\qquad 0\le\delta\le1.
\]
This is a feasible \emph{candidate}, not an assumption restricting the seller's experiment space. After either signal, optimal verification in the unique symmetric equilibrium is
\[
e^*(\delta)=\frac{1-\delta}{4\kappa+1-\delta}.
\]
Directly evaluating the second-highest-bid distribution at this equilibrium gives
\[
\Pi(\delta;\kappa)=\frac{\delta^2-16\delta\kappa^2-8\delta\kappa-2\delta+32\kappa^2+8\kappa+1}{4(4\kappa+1-\delta)^2},
\qquad
\frac{\partial\Pi}{\partial\delta}=\frac{4\kappa^2(1-4\kappa-\delta)}{(4\kappa+1-\delta)^3}.
\]
Set $\kappa=1/100$. The derivative is positive for $\delta<24/25$ and negative for $\delta>24/25$. Thus the feasible policy $\delta=24/25$ strictly beats both endpoints $\delta=0$ (no disclosure) and $\delta=1$ (full disclosure). The benchmark disclosure structure is therefore not optimal.

\emph{Cases $N=3,4$.} Given posterior $p$, rival-bid cdf $H$, and $\Phi_H(x)=\int_0^x H(t)^{N-1}\,dt$, the acquisition value is
\[
 A(p,H)=p\Phi_H(1)-\Phi_H(p).
\]
Since $0\le H^{N-1}\le1$ and $\Phi_H$ is convex, $0\le A(p,H)\le1$. Quadratic verification costs imply
\[
0\le e(p,H)=\min\{1,A(p,H)/\kappa\}\le1/\kappa.
\]
Coupling the realized bids with the no-verification bids, they can differ only if at least one bidder verifies; the union bound gives probability at most $N/\kappa$. The second-highest bid always lies in $[0,1]$. Accordingly, for any feasible experiment,
\begin{equation}\label{eq:binary-coupling-bound}
\left|R_N^{\rm verify}(\nu;\kappa)-R_N^{\rm no\ verify}(\nu)\right|\le N/\kappa.
\end{equation}
For three bidders use
\[
\nu_3=\frac14\delta_0+\frac34\delta_{2/3}.
\]
Its mean is $1/2$, so it is Bayes-plausible. It can be implemented by $\Pr(s=L\mid v=0)=1/2$ and $\Pr(s=L\mid v=1)=0$. The no-verification second-highest bid equals $2/3$ if at least two bidders receive the high signal. Consequently,
\[
R_3^{\rm no\ verify}(\nu_3)=\frac23\left[3\left(\frac34\right)^2\frac14+\left(\frac34\right)^3\right]=\frac9{16}.
\]
Full disclosure yields exactly $1/2$ at any $\kappa$, since degenerate posteriors have zero verification value. Full pooling yields at most $1/2+3/\kappa$ by \eqref{eq:binary-coupling-bound}. The proposed experiment yields at least $9/16-3/\kappa$, which exceeds both at $\kappa=100$. Hence neither benchmark disclosure policy is optimal for $N=3$.

For four bidders, use the asymmetric feasible posterior distribution
\[
\nu_4=\frac7{15}\delta_0+\frac8{15}\delta_{15/16},
\]
implemented by $\Pr(s=L\mid v=0)=14/15$ and $\Pr(s=L\mid v=1)=0$. Its mean is $1/2$, and direct binomial calculation gives
\[
R_4^{\rm no\ verify}(\nu_4)=\frac{15}{16}\left[1-\left(\frac7{15}\right)^4-4\frac8{15}\left(\frac7{15}\right)^3\right]=\frac{776}{1125}.
\]
Full disclosure yields $11/16$. The revenue advantage in the no-verification limit is
\[
\frac{776}{1125}-\frac{11}{16}=\frac{41}{18000}>0.
\]
By \eqref{eq:binary-coupling-bound}, the verified revenue under $\nu_4$ exceeds full revelation when $\kappa>4/(41/18000)=72000/41$. At $\kappa=2000$ it also exceeds full pooling, whose revenue is at most $1/2+4/\kappa$. Hence neither benchmark disclosure policy is optimal for $N=4$.

\emph{Case $N\ge5$.} Write the revenue from any equilibrium final-bid cdf $H$ as
\[
R_N(H)=\int_0^1 f_N(H(t))\,dt,\qquad f_N(x)=1-Nx^{N-1}+(N-1)x^N.
\]
Every final bid equals a conditional expectation of the true valuation; hence it has mean $1/2$, so $\int_0^1H(t)\,dt=1/2$. We show
\begin{equation}\label{eq:binary-tangent}
f_N(x)\le f_N(1/2)+f_N'(1/2)(x-1/2),\quad 0\le x\le1.
\end{equation}
Indeed,
\[
f_N''(x)=-N(N-1)x^{N-3}\big[(N-2)-(N-1)x\big].
\]
Thus $f_N$ is concave on $[0,(N-2)/(N-1)]$ and convex thereafter; the inflection point exceeds $1/2$. The tangent at $1/2$ lies above $f_N$ on the entire concave interval. On the remaining interval the difference between that tangent and $f_N$ is concave, so its minimum is at an endpoint. It is nonnegative at the inflection point, and at $x=1$ equals
\[
 f_N(1/2)+\tfrac12f_N'(1/2)-f_N(1)=1-\frac{N^2+1}{2^N}\ge0 \quad(N\ge5).
\]
Here $2^N\ge N^2+1$ at $N=5$ and is preserved inductively. This proves \eqref{eq:binary-tangent}. Integrating and using the mean constraint yields
\[
 R_N(H)\le f_N(1/2)=1-\frac{N+1}{2^N}.
\]
Under full disclosure the final bids equal the independent binary valuations, $H(t)=1/2$ for $0\le t<1$, attaining this bound. Moreover, the tangent bound is strict unless $H(t)=1/2$, so attaining equality requires the final bids to have the binary distribution $\frac12\delta_0+\frac12\delta_1$. Under any nondegenerate initial posterior $p\in(0,1)$, this would require verification with probability one. Yet against the binary final-bid distribution, $\Phi_H$ is affine on $[0,1]$ and $A(p,H)=p\Phi_H(1)-\Phi_H(p)=0$; strictly costly verification therefore cannot occur. Consequently, full disclosure is the unique optimal disclosure policy. Its explicit rule is
\[
s_i^*(v_i)=v_i,\qquad v_i\in\{0,1\},\quad N\ge5.
\]
Therefore, a policy with a nonempty high-value pooling region cannot be optimal.
\end{proof}

Taken together, the uniform and binary examples show that endogenous costly private verification can lead to disclosure structures substantially different from full revelation for low values and single pooling for high values, even for two of the most familiar valuation distributions. Section~5 develops the general characterization of optimal disclosure.

\begin{remark}[Reserve prices and alternative auction formats]
\label{rem:auction-extensions}
The preceding results qualify a particularly strong benchmark finding of Bergemann et al. (2022): their policy of revealing low values while pooling high values has a critical quantile that depends on the number of bidders but not on the valuation distribution. With endogenous costly verification, this policy need not remain revenue-maximizing even for familiar uniform and binary valuations. Two variations considered in the auction literature clarify the scope of this economic mechanism. First, a reserve price changes the range of bids that can win and hence the expected return to private verification. When the seller chooses a reserve price along with a disclosure experiment, the reserve and the induced Bayesian Nash verification choices may need to be optimized jointly. The informational effect identified above remains relevant, but the optimal reserve and the optimal disclosure structure do not follow from the zero-reserve results without a new equilibrium and revenue analysis.

Second, BHMSW use revenue equivalence to relate their second-price benchmark to other classical auction formats. Other pricing and allocation rules, including uniform-price auctions, can change how a bidder's information affects winning, payments, and competition, thereby changing the private value of verification. Optimal disclosure under these auction formats consequently depends on the corresponding verification equilibria and revenue effects.
\end{remark}

The failure of the benchmark disclosure policy to remain generally optimal is most naturally read against the established literature on information in auctions \citep{MilgromWeber1982Information,BergemannPesendorfer2007,BHMSW2022}. The economic departure is that initial seller information changes the private information acquisition game rather than merely changing posterior bids. Section~5 studies the equilibrium-implementability consequences of this departure.

\section{Revenue Decomposition and General Optimality}
\label{sec:general-characterization}

Section~3 establishes the symmetric verification equilibrium for an arbitrary Bayes-plausible experiment, and Section~4 shows that the familiar benchmark pattern of revealing lower valuations and pooling higher valuations need not remain optimal when verification is endogenous. The purpose of this section is to develop the general theory behind that finding. We first decompose seller revenue into information and verification-related components. We then characterize the final-bid distributions that can arise in verification equilibrium and state exact necessary and sufficient conditions for global optimality. Finally, we identify the cross-signal incentive interaction that makes disclosure and verification inseparable in general. The three parts are complementary: revenue decomposition identifies the objective, implementability restricts the attainable outcomes, and incentive interaction explains why those restrictions respond to the seller's disclosure choice.

The argument draws on the information-design formulation of Bergemann, Heumann, Morris, Sorokin, and Winter (2022) while keeping endogenous acquisition incentives explicit. Unlike the general dynamic screening framework, the auction studied here is static; the relevant interaction is captured by the equilibrium verification system rather than intertemporal incentive constraints.

\subsection{Revenue Decomposition and Economic Trade-offs}
\label{subsec:two-tradeoffs}

The seller's objective is revenue, not an information order or a verification potential in isolation. Before studying which outcomes are implementable, we identify the economic forces that the seller must balance. Initial disclosure can compress bids and save bidders' competitive rents, but may sacrifice allocative efficiency. Endogenous verification partially restores information while changing the competitive environment and giving bidders both incentives and rents from subsequent information acquisition. The following decomposition isolates these two interdependent trade-offs for every Bayes-plausible experiment, without assuming that optimal disclosure has a particular structure.

Fix any Bayes-plausible experiment $\pi\in\Pi_F$, and write $\mu$ for a bidder's posterior and $m=m(\mu)$ for its mean. Let $G^0_\pi$ denote the distribution of $m$ under $\pi$; it is the final-bid distribution that would arise if verification were unavailable. Let $G^p_\pi$ denote the actual symmetric-equilibrium final-bid distribution, and abbreviate it by $H$. The corresponding optimal verification probability is $e(\mu,H)$. Thus
\[
 G^0_\pi=\int_K\delta_{m(\mu)}\,\pi(d\mu),\qquad
 H=\int_K\bigl[(1-e(\mu,H))\delta_{m(\mu)}+e(\mu,H)\mu\bigr]\,\pi(d\mu).
\]
Both $G^0_\pi$ and $H$ are mean-preserving contractions of the prior $F$, but the former depends only on disclosure whereas the latter also incorporates endogenous verification.

For a probability distribution $J$ on $V=[0,M]$, define its expected highest bid and total bidder gross surplus by
\[
 \mathsf W_N(J)=\mathbb E_{J^{\otimes N}}\bigl[\max\{B_1,\ldots,B_N\}\bigr],\qquad
 \mathsf B_N(J)=\mathsf W_N(J)-R_N(J).
\]
The second quantity is the expected sum of the differences between the highest and second-highest bids. It is nonnegative and equals
\[
 \mathsf B_N(J)=N\int_V\Phi_J(b)\,J(db),\qquad
 \Phi_J(v)=\int_0^v J(t)^{N-1}\,dt.
\]
These identities remain valid when $J$ has atoms. For a posterior $\mu$ and a candidate competitor distribution $J$, let
\[
 A(\mu,J)=\int_V\Phi_J(v)\,\mu(dv)-\Phi_J(m(\mu)).
\]
Convexity of $\Phi_J$ implies $A(\mu,J)\geq0$. In equilibrium, the bidder selects an effort maximizing $eA(\mu,H)-c(e)$ over $0\leq e\leq1$.

\begin{proposition}[Exact revenue decomposition and two disclosure trade-offs]
\label{prop:two-tradeoff-decomposition}
Under the maintained assumptions of Sections 2 and 3, every Bayes-plausible experiment $\pi$ and its symmetric verification equilibrium satisfy
\begin{equation}
\begin{aligned}
R_N(H)-R_N(F)
={}&
\underbrace{\mathsf B_N(F)-\mathsf B_N(G^0_\pi)}_{\text{rents saved through initial disclosure}}
-\underbrace{[\mathsf W_N(F)-\mathsf W_N(H)]}_{\text{remaining allocation loss}}\\
&-\underbrace{N\,\mathbb E_\pi[\Phi_H(m)-\Phi_{G^0_\pi}(m)]}_{\text{verification-induced competitive-rent adjustment}}
-\underbrace{N\,\mathbb E_\pi[e(\mu,H)A(\mu,H)]}_{\text{verification rent and resource-cost wedge}}.
\end{aligned}
\label{eq:two-tradeoff-decomposition}
\end{equation}
Equivalently, the revenue change can be separated into the classical initial-disclosure effect and the endogenous-verification effect:
\begin{equation}
\begin{aligned}
 R_N(H)-R_N(F)
 ={}&\underbrace{\bigl[\mathsf B_N(F)-\mathsf B_N(G^0_\pi)\bigr]
       -\bigl[\mathsf W_N(F)-\mathsf W_N(G^0_\pi)\bigr]}_{R_N(G^0_\pi)-R_N(F)}\\
 &+\underbrace{\bigl[\mathsf W_N(H)-\mathsf W_N(G^0_\pi)\bigr]
  -N\mathbb E_\pi[\Phi_H(m)-\Phi_{G^0_\pi}(m)]
  -N\mathbb E_\pi[e(\mu,H)A(\mu,H)]}_{R_N(H)-R_N(G^0_\pi)}.
\end{aligned}
\label{eq:two-tradeoff-benchmark}
\end{equation}
The allocation terms have the unambiguous signs
\[
 \mathsf W_N(F)\geq\mathsf W_N(H)\geq\mathsf W_N(G^0_\pi).
\]
The competitive-rent adjustment in \eqref{eq:two-tradeoff-decomposition} need not have a fixed sign.
\end{proposition}

\begin{proof}
For any bid distribution $J$, the highest bid is the second-highest bid plus the sum of the bidders' ex post gross bidding surpluses. Taking expectations gives $R_N(J)=\mathsf W_N(J)-\mathsf B_N(J)$ and $\mathsf B_N(J)=N\int\Phi_J\,dJ$. In particular, using the expression for the equilibrium bid law,
\[
\begin{aligned}
\mathsf B_N(H)
 &=N\mathbb E_\pi\bigl[(1-e)\Phi_H(m)+e\textstyle\int\Phi_H(v)\,\mu(dv)\bigr]\\
 &=N\mathbb E_\pi[\Phi_H(m)]+N\mathbb E_\pi[eA(\mu,H)].
\end{aligned}
\]
As $\mathsf B_N(G^0_\pi)=N\mathbb E_\pi[\Phi_{G^0_\pi}(m)]$, subtracting the two gross-surplus expressions yields
\[
\mathsf B_N(H)-\mathsf B_N(G^0_\pi)
 =N\mathbb E_\pi[\Phi_H(m)-\Phi_{G^0_\pi}(m)]
  +N\mathbb E_\pi[eA(\mu,H)].
\]
Add and subtract $R_N(G^0_\pi)=\mathsf W_N(G^0_\pi)-\mathsf B_N(G^0_\pi)$ to obtain \eqref{eq:two-tradeoff-benchmark}, and rearrange the allocation terms to obtain \eqref{eq:two-tradeoff-decomposition}.

For the allocation inequalities, couple a bidder's true value $v$, posterior mean $m$, and realized final bid $B$. Conditional on the initial signal, $\mathbb E[v]=m$; conditional on the final information, $\mathbb E[v]=B$. Thus $G^0_\pi\preceq_{\mathrm{cx}}H\preceq_{\mathrm{cx}}F$. Independent copies of these couplings and convexity of the maximum in its arguments imply the displayed inequalities. No ordering of the remaining rent-adjustment terms follows from convex order.
\end{proof}

The first bracket in \eqref{eq:two-tradeoff-benchmark} is the familiar information-design trade-off. Pooling redistributes the information available to bidders and can reduce their aggregate bidding surplus, which benefits the seller, but it also sacrifices allocative efficiency. The second bracket isolates the additional trade-off created by privately chosen verification. Verification restores some allocative efficiency, but changes the competitive environment faced by every posterior type and allows bidders to retain part of the returns from verification. It is therefore not correct to regard hidden verification as simply adding an exogenous implementation cost to the classical objective. In particular, an experiment maximizing revenue when verification is unavailable need not maximize revenue when verification is endogenous.

The last term has a precise rent-and-cost interpretation. Define the constrained convex conjugate of the verification cost by
\[
 c^*(a)=\max_{0\leq z\leq1}\{za-c(z)\},\qquad a\geq0.
\]
At an optimal verification choice, $eA=c(e)+c^*(A)$. The first term is a real information-acquisition cost and the second is the bidder's additional \emph{net} payoff from the option to verify. Hence both costs and incentive rents enter the seller's revenue accounting, but neither is deducted twice. For the quadratic benchmark $c(e)=\kappa e^2/2$, the combined wedge satisfies
\begin{equation}
 eA=\begin{cases}
 A^2/\kappa,&0\leq A\leq\kappa,\\
 A,&A\geq\kappa.
 \end{cases}
\label{eq:verification-wedge-quadratic}
\end{equation}
The saturation branch matters: the interior quadratic expression cannot be extended to signals that induce verification with probability one.

The decomposition can also be written against the full-information efficient allocation. Indeed, the actual bidder's expected \emph{net} utility after observing posterior $\mu$ is
\[
 U(\mu,H)=\Phi_H(m)+c^*(A(\mu,H)).
\]
Transfers cancel between the seller and the bidders, while the verification cost is a real resource loss. Thus Proposition \ref{prop:two-tradeoff-decomposition} is equivalent to
\begin{equation}
 R_N(H)=\mathsf W_N(F)-\bigl[\mathsf W_N(F)-\mathsf W_N(H)\bigr]
 -N\mathbb E_\pi[U(\mu,H)]-N\mathbb E_\pi[c(e)].
\label{eq:revenue-surplus-rent-cost}
\end{equation}
Equation \eqref{eq:revenue-surplus-rent-cost} is the static counterpart of the surplus--information-rent--implementation-cost accounting familiar from screening. Its economic content is exact, but it is not a Myerson-type virtual-surplus reduction: bidders do not observe and report their original values before disclosure, and no truthful-reporting envelope justifies replacing their posterior-dependent rents with the classical inverse-hazard-rate adjustment.

\begin{proposition}[Cost comparative statics of equilibrium verification]
\label{prop:verification-cost-comparative-statics}
Fix a Bayes-plausible experiment $\pi$ and suppose verification costs are $c_\kappa(e)=\kappa e^2/2$ for $\kappa>0$. Let $e_\kappa$ be the unique symmetric equilibrium verification rule, and write $G_e^p$ for the final-bid law generated by $\pi$ and an arbitrary feasible effort rule $e$. Define
\[
 \mathcal V_\pi(\kappa)
 =\min_{0\leq e\leq1}\left\{
 \frac1N\int_0^M G_e^p(b)^N\,db
 +\frac\kappa2\int_K e(\mu)^2\,\pi(d\mu)
 \right\}.
\]
Then $\mathcal V_\pi$ is nondecreasing and concave in $\kappa$. Wherever it is differentiable,
\[
 \mathcal V_\pi'(\kappa)=\frac12\mathbb E_\pi[e_\kappa^2].
\]
In particular, whenever $0<\kappa_1<\kappa_2$,
\[
 \mathbb E_\pi[e_{\kappa_2}^2]\leq\mathbb E_\pi[e_{\kappa_1}^2].
\]
This comparison does not generally order equilibrium revenue or verification effort separately at every posterior.
\end{proposition}

\begin{proof}
For fixed $\pi$ and $e$, the expression minimized in the definition of $\mathcal V_\pi$ is affine and nondecreasing in $\kappa$. Its pointwise infimum is therefore nondecreasing and concave. Section 3's convex-potential result identifies its unique minimizer with the symmetric verification equilibrium. The envelope formula follows from the envelope theorem (see \citealp{MilgromSegal2002,BonnansShapiro2000}); uniqueness also yields the stated one-sided derivative comparison when needed. Since derivatives of a concave function are nonincreasing, the expected squared verification effort is nonincreasing in the cost coefficient. The result concerns the aggregate squared effort, not posterior-by-posterior monotonicity of effort or monotonicity of seller revenue.
\end{proof}

The direction of the revenue effect can be seen in the uniform, three-bidder interval-pooling benchmark derived in the interval-pooling analysis below. Pooling an interval $[a,b]$, with width $d=b-a$ and mean $m=(a+b)/2$, produces the gain
\[
 W_\kappa(a,b)=\frac{(a+b-1)d^3}{4}(1-e_\kappa)(2-e_\kappa),\qquad
 e_\kappa=\min\left\{1,\frac{3md^2}{12\kappa+2md^2}\right\}.
\]
For $e_\kappa<1$, effort decreases with $\kappa$. Hence increasing verification costs increases the revenue contribution of an \emph{upper} pooling interval ($a+b>1$), but decreases the contribution of a \emph{lower} pooling interval ($a+b<1$); when $a+b=1$, the contribution vanishes. This does not rank the seller's unrestricted optimum across cost parameters, because the seller also reoptimizes the disclosure experiment. It does exhibit the two opposing forces explicitly: more expensive verification preserves the rent-saving consequences of upper pooling while limiting the allocative information subsequently recovered by bidders.

The accounting identity therefore supplies the objective for the unrestricted implementation analysis that follows. Writing $G^{p,*}(\pi)$ for the unique verification-equilibrium bid distribution, the seller's value can be expressed as
\begin{equation}
\begin{aligned}
 \sup_{\pi\in\Pi_F}R_N(G^{p,*}(\pi))
 =R_N(F)+\sup_{\pi\in\Pi_F}\Big\{&R_N(G^0_\pi)-R_N(F)\\
 &+R_N(G^{p,*}(\pi))-R_N(G^0_\pi)\Big\}.
\end{aligned}
\label{eq:joint-two-tradeoff-program}
\end{equation}
The two effects cannot be maximized separately: each depends on the same disclosure experiment and the equilibrium verification response. This decomposition identifies precisely what any global seller optimum must balance, but it does not by itself establish which equilibrium final-bid distributions are attainable. That implementation question is taken up next, before we turn to the reduced optimality conditions.

\subsection{Equilibrium Implementability and Global Characterization of Optimal Disclosure}
\label{subsec:equilibrium-implementability}

The revenue decomposition identifies the economic objective, but a seller cannot arbitrarily specify the final-bid distribution that enters it. Given an initial experiment, bidders choose verification optimally, and these decisions jointly determine the distribution of bids that a mechanism implements. We therefore turn from revenue accounting to the unrestricted feasibility and global optimality problem, without restricting posterior supports. For any final-bid distribution $G^p$, expected revenue is the expected second-highest bid, as established in Section 2:
\begin{equation}
R_N(G^p)=\int_0^M\left[1-N[G^p(b)]^{N-1}+(N-1)[G^p(b)]^N\right]db.
\label{eq:revenue-unrestricted}
\end{equation}
The seller chooses $\pi\in\Pi_F$ to maximize $R_N(G^{p,*}(\pi))$, where $G^{p,*}(\pi)$ is the unique symmetric equilibrium final-bid distribution characterized in Theorem~\ref{thm:cont}. Theorem~\ref{thm:seller} guarantees existence of an optimal experiment, but neither identifies which bid distributions are implementable nor specifies their optimal structure.

In the classical model, a distribution of posterior expected valuations is feasible if and only if it is a mean-preserving contraction of $F$. With endogenous verification, the final-bid distribution must satisfy the corresponding information order, but this condition is no longer sufficient. Information acquired after disclosure is determined by bidders' privately optimal decisions, not selected by the seller. A final-bid distribution can therefore respect Bayes plausibility without arising in any symmetric verification equilibrium.

Fix a candidate final-bid distribution $G^p$. For each posterior $\mu\in K$, Section 3 defines the verification best response $e(\mu,G^p)$ and the resulting conditional final-bid distribution $T_{G^p}(\mu)$ in \eqref{eq:BestResponseGeneral}. The set of distributions produced by optimal verification responses to the fixed candidate $G^p$ is
\[
\mathcal C_F(G^p)=\left\{\int_K T_{G^p}(\mu)\,\pi(d\mu):\pi\in\Pi_F\right\}.
\]
Holding $G^p$ fixed, the map from experiments to the resulting distributions is affine and continuous. As $\Pi_F$ is compact and convex, so is $\mathcal C_F(G^p)$.

A candidate final-bid distribution is implementable only if some Bayes-plausible experiment induces it through the selected symmetric verification equilibrium of Definition~\ref{def:verification-bne}. Since $\mathcal C_F(G^p)$ already incorporates the optimal verification response to the fixed candidate $G^p$, this requirement is precisely a fixed-point condition.

\begin{definition}[Equilibrium-Implementable Bid Distribution]
\label{def:equilibrium-implementable-bids}
A distribution $G^p\in\mathcal P(V)$ is \emph{equilibrium implementable} if and only if
\[
G^p\in\mathcal C_F(G^p).
\]
\end{definition}
The distinction between $\mathcal C_F(G^p)$ and the entire equilibrium-implementable set matters. The former is convex because $G^p$ is held fixed. The latter collects distributions that belong to their own corresponding sets $\mathcal C_F(G^p)$, and need not be convex. Endogenous verification therefore prevents the seller's problem from becoming a conventional convex optimization over final-bid distributions.

We express this fixed-point requirement as inequalities. Continuous functions on the bid space assign hypothetical values to final bids and, through separation, test whether a proposed bid distribution belongs to $\mathcal C_F(G^p)$. Let $C(V)$ denote the continuous real-valued functions on $V$. Given $G^p$ and $g\in C(V)$, define
\begin{equation}
\begin{split}
h_{G^p,g}(\mu)&=\int_V g(b)\,T_{G^p}(\mu)(db),\\
\operatorname{cav}h_{G^p,g}(F)&=\max_{\pi\in\Pi_F}\int_K h_{G^p,g}(\mu)\,\pi(d\mu).
\end{split}
\label{eq:posterior-test-payoff}
\end{equation}
The first expression is the expected test payoff when a bidder starts from posterior $\mu$ and verifies optimally against $G^p$. The second is the largest attainable expected test payoff among all Bayes-plausible experiments. This is a concavification of a test payoff, not a concavification of seller revenue.

\begin{theorem}[Exact Test-Function and Posterior-Price Characterization]
\label{thm:exact-test-posterior-price}
Under the assumptions of Theorem~\ref{thm:cont}, a distribution $G^p\in\mathcal P(V)$ can arise as the symmetric equilibrium final-bid distribution for some Bayes-plausible private disclosure experiment if and only if
\begin{equation}
\int_V g\,dG^p\leq\operatorname{cav}h_{G^p,g}(F)
\qquad\text{for every }g\in C(V).
\label{eq:implementability-test}
\end{equation}
Moreover, the concavification value has the posterior-price dual representation
\begin{equation}
\operatorname{cav}h_{G^p,g}(F)
=\inf_{\varphi\in C(V)}\left\{\int_V\varphi(u)\,F(du):
\int_V\varphi(u)\,\mu(du)\geq h_{G^p,g}(\mu)
\text{ for every }\mu\in K\right\}.
\label{eq:posterior-price-dual}
\end{equation}
The infimum in \eqref{eq:posterior-price-dual} need not be attained without an additional supporting-price regularity condition.
\end{theorem}

\begin{proof}
Fix $G^p$. By Theorem~\ref{thm:cont}, the optimal verification effort at every posterior is uniquely determined, and $\mu\mapsto T_{G^p}(\mu)$ is weakly continuous. The mapping
\[
\pi\longmapsto\int_K T_{G^p}(\mu)\,\pi(d\mu)
\]
is continuous and affine. Since $\Pi_F$ is compact and convex, $\mathcal C_F(G^p)$ is a compact convex subset of the signed measures on $V$ endowed with the weak topology.

Membership in this set is equivalent to
\[
\int_V g\,dG^p\leq \sup_{H\in\mathcal C_F(G^p)}\int_V g\,dH
\qquad\text{for every }g\in C(V).
\]
Necessity is immediate. Conversely, if $G^p\notin\mathcal C_F(G^p)$, strict separation of a point from a compact convex set supplies a weakly continuous linear functional separating them. Every such functional is integration against some $g\in C(V)$, contradicting the inequalities. By definition,
\[
\sup_{H\in\mathcal C_F(G^p)}\int_V g\,dH
=\max_{\pi\in\Pi_F}\int_K\left[\int_V g(b)\,T_{G^p}(\mu)(db)\right]\pi(d\mu)
=\operatorname{cav}h_{G^p,g}(F).
\]
Thus \eqref{eq:implementability-test} is equivalent to $G^p\in\mathcal C_F(G^p)$, which is exactly equilibrium implementability.

For the dual, $h_{G^p,g}$ is continuous on the compact convex space $K=\mathcal P(V)$. Every continuous affine functional on $K$ has the form $\mu\mapsto\int_V\varphi\,d\mu$ for some $\varphi\in C(V)$. Any such functional majorizing $h_{G^p,g}$ satisfies, for $\pi\in\Pi_F$,
\[
\int_K h_{G^p,g}(\mu)\,\pi(d\mu)
\leq\int_K\int_V\varphi(u)\,\mu(du)\,\pi(d\mu)
=\int_V\varphi(u)\,F(du).
\]
Hence the right-hand side of \eqref{eq:posterior-price-dual} bounds the concavification value from above. Conversely, let $C(F)=\max_{\pi\in\Pi_F}\int_Kh_{G^p,g}\,d\pi$.
For each $\varepsilon>0$, the hypograph of the concave envelope of
$h_{G^p,g}$ can be strictly separated from the point $(F,C(F)+\varepsilon)$
in the locally convex space of signed measures crossed with $\mathbb R$.
The separation yields a continuous affine majorant $L$ of
$h_{G^p,g}$ with $L(F)<C(F)+\varepsilon$; the no-duality-gap result of \citet[Theorem~2]{DworczakKolotilin2024}
provides this representation for bounded upper-semicontinuous posterior
objectives on compact state spaces, as satisfied here by continuity of
$h_{G^p,g}$. Since continuous
affine functions on $K=\mathcal P(V)$ have the form
$L(\mu)=\int_V\varphi\,d\mu$ for some $\varphi\in C(V)$,
this produces a feasible price with objective below $C(F)+\varepsilon$.
Sending $\varepsilon\downarrow0$ establishes \eqref{eq:posterior-price-dual}, without asserting
dual attainment.
\end{proof}

Theorem \ref{thm:exact-test-posterior-price} gives exact implementability conditions for arbitrary posterior alphabets. They strengthen the mean-preserving-spread restriction by imposing verification incentives. The posterior-price representation expresses the same conditional linear design problem in terms of a price $\varphi(u)$ assigned to underlying valuations: each posterior's expected price must cover the test payoff generated by optimal verification. Bayes plausibility then turns that pointwise majorization into an aggregate bound evaluated at $F$.

\begin{proposition}[Conditional Posterior Contact]
\label{prop:conditional-posterior-contact}
Fix $G^p$ and $g\in C(V)$. Suppose the infimum in \eqref{eq:posterior-price-dual} is attained at $\varphi^*$, and let $\pi^*\in\arg\max_{\pi\in\Pi_F}\int_K h_{G^p,g}(\mu)\,\pi(d\mu)$. Then $\pi^*$ is supported on the contact set
\begin{equation}
\mathcal A_{G^p,g,\varphi^*}
=\left\{\mu\in K:\int_V\varphi^*(u)\,\mu(du)=h_{G^p,g}(\mu)\right\}.
\label{eq:posterior-contact-set}
\end{equation}
In particular, if this contact set is finite, every maximizing experiment for this fixed-$G^p$, fixed-$g$ concavification has finite support. This conclusion does not establish finite support for a globally optimal seller experiment.
\end{proposition}

\begin{proof}
Primal and dual optimality, together with \eqref{eq:posterior-price-dual}, imply
\[
\int_K h_{G^p,g}(\mu)\,\pi^*(d\mu)=\int_V\varphi^*(u)\,F(du).
\]
By Bayes plausibility the latter integral equals
\[
\int_K\left[\int_V\varphi^*(u)\,\mu(du)\right]\pi^*(d\mu).
\]
Therefore
\[
\int_K\left[\int_V\varphi^*(u)\,\mu(du)-h_{G^p,g}(\mu)\right]\pi^*(d\mu)=0.
\]
The integrand is continuous and nonnegative by dual feasibility and hence vanishes $\pi^*$-almost surely. This is precisely the asserted contact condition. If the contact set is finite, the support of any maximizing experiment is contained in that finite set.
\end{proof}

Posterior contact is a powerful restriction for a specified linear concavification problem, but it is conditional on both the rival-bid distribution and the test payoff. Seller revenue is nonlinear in the endogenous equilibrium bid distribution. Consequently, \eqref{eq:posterior-contact-set} alone is not a characterization of revenue-maximizing pooling regions.

We now return to the seller's actual objective. By Theorem \ref{thm:exact-test-posterior-price}, the distributions satisfying the test-function inequalities are exactly the symmetric equilibrium-implementable final-bid distributions. This leads to a criterion for global seller optimality which does not presume that revenue is concave.

\begin{proposition}[Global Optimality over Equilibrium-Implementable Distributions]
\label{prop:global-equilibrium-optimality}
Let $\pi^*\in\Pi_F$ induce the equilibrium final-bid distribution $G^{p,*}=G^{p,*}(\pi^*)$. The experiment $\pi^*$ is globally optimal for the seller if and only if
\begin{equation}
R_N(H)\leq R_N(G^{p,*})
\quad\text{for every }H\in\mathcal P(V)\text{ with }H\in\mathcal C_F(H).
\label{eq:unrestricted-global-optimality}
\end{equation}
A sufficient, generally stronger, certificate is the existence of $g^*\in C(V)$ such that, for every equilibrium-implementable $H$,
\begin{equation}
\begin{aligned}
R_N(H)-R_N(G^{p,*})&\leq\int_V g^*\,d(H-G^{p,*}),\\
\int_V g^*\,dH&\leq\int_V g^*\,dG^{p,*}.
\end{aligned}
\label{eq:global-support-certificate}
\end{equation}
Condition \eqref{eq:unrestricted-global-optimality} is necessary and sufficient. The certificate \eqref{eq:global-support-certificate} is sufficient, but it need not exist for every global optimum.
\end{proposition}

\begin{proof}
By Theorem~\ref{thm:cont} and Theorem \ref{thm:exact-test-posterior-price}, the distributions satisfying \eqref{eq:implementability-test} are precisely those induced by symmetric verification equilibria under Bayes-plausible experiments. The experiment $\pi^*$ therefore maximizes seller revenue across all admissible experiments if and only if no such implementable $H$ yields greater revenue, which is \eqref{eq:unrestricted-global-optimality}.

If \eqref{eq:global-support-certificate} holds, then for every implementable $H$,
\[
R_N(H)-R_N(G^{p,*})\leq \int_V g^*\,d(H-G^{p,*})\leq0.
\]
Thus \eqref{eq:unrestricted-global-optimality} follows and $\pi^*$ is globally optimal.
\end{proof}

Proposition \ref{prop:global-equilibrium-optimality} completes the unrestricted implementability and optimality characterization. Its first condition is an exact global test, but not an explicit solution in terms of $F$ and $c$; the test still ranges over an infinite-dimensional implementable set. The second gives a potentially more tractable supporting certificate, provided both inequalities can actually be proved. Neither follows merely from convexity of the verification potential.

These results exactly characterize the unrestricted problem, but their test-function formulation need not make a candidate policy easy to verify. Before turning to those reductions, we isolate the incentive interaction underlying the equilibrium implementability restrictions. Section~6 will then use the characterized equilibrium to develop sharper necessary conditions and comparative statics.

The distinction between selecting a signal structure and inducing the information ultimately used in decisions is closely related to joint information and incentive design \citep{BergemannHeumannMorris2026,BergemannHeumannMorrisJPE2026}. Related dynamic information environments highlight the informational consequences of feedback and incentive choices \citep{ElyGeorgiadisRayo2025,Tian2026InformationalEvolution}. The present nonseparability is generated by the bidders' static Bayesian Nash verification choices, and does not rely on an intertemporal contracting interpretation.

\subsection{Disclosure--Verification Nonseparability}
\label{subsec:disclosure-verification-nonseparability}

The global characterization in Subsection~\ref{subsec:equilibrium-implementability} treats the final-bid distribution as an endogenous equilibrium object. This qualification is economically substantive: a change in the seller's disclosure experiment alters the value of private verification, while verification changes the distribution of rival bids that determines the value of information after every other signal. We next identify this interaction directly from the verification equilibrium. The argument is static and uses the convex-potential structure of Section~3, rather than any dynamic incentive or recursive formulation.

Consider a finite experiment with positive signal probabilities $\pi_s$, posterior distributions $\mu_s$, means $m_s$, and verification probabilities $e_s$. Write $H_s$ for the cdf of $\mu_s$ and let
\[
 \Delta_s(b)=H_s(b)-\mathbf 1\{m_s\le b\},\qquad
 G^{p}_e(b)=\sum_s\pi_s\bigl[\mathbf 1\{m_s\le b\}+e_s\Delta_s(b)\bigr].
\]
The value of verification after signal $s$ can be expressed, by the definitions in Section~3 and an integration-by-parts identity, as
\[
 A_s(G^{p}_e)=-\int_0^M G^{p}_e(b)^{N-1}\Delta_s(b)\,db.
\]
The dependence of $A_s$ on other signals' verification efforts provides an exact measure of their interaction.

\begin{proposition}[Cross-signal verification incentives]
\label{prop:cross-signal-verification}
Fix a finite experiment and an effort profile $e\in[0,1]^S$. For any signals $s,r$,
\begin{equation}\label{eq:cross-verification}
 \frac{\partial A_s(G^{p}_e)}{\partial e_r}
 =-(N-1)\pi_r\int_0^M G^{p}_e(b)^{N-2}\Delta_s(b)\Delta_r(b)\,db.
\end{equation}
Consequently, the marginal private value of verifying after signal $s$ is locally independent of effort after signal $r$ precisely when the weighted integral in \eqref{eq:cross-verification} vanishes. In particular, the cross effect vanishes when $\Delta_s\Delta_r=0$ almost everywhere. If $m_s=m_r$ and the weighted product is strictly positive on a set of positive measure, then $\partial A_s/\partial e_r<0$: the two verification margins are strategic substitutes at that effort profile.
\end{proposition}
\begin{proof}
For fixed posterior distributions and means, $\partial G^{p}_e(b)/\partial e_r=\pi_r\Delta_r(b)$. Differentiating the integral representation of $A_s$ under the integral sign, justified by boundedness on the compact bid interval, gives \eqref{eq:cross-verification}. If $m_s=m_r=m$, both $\Delta_s$ and $\Delta_r$ are nonnegative below $m$ and nonpositive above $m$, so their product is nonnegative almost everywhere; strict positivity of its weighted integral yields the claimed negative sign. The zero-effect claim follows directly from the same integral.
\end{proof}

The last conclusion is stronger than observing that a single bidder's verification choice depends on her posterior: it identifies a cross-signal competitive externality, even though bidders' underlying values are independent. The terminology of affiliation used for correlated types in dynamic screening is not needed here. Common posterior means furnish a particularly transparent sufficient condition for a signed interaction; arbitrary pairs of posteriors need not have a determinate cross-effect sign.

For an interior equilibrium along a differentiable finite-dimensional family of Bayes-plausible experiments $t\mapsto\pi_t$, let $\mathcal B_s(e,t)=c'(e_s)-A_s(G^p_{e,t})$ and write $J=D_e\mathcal B$. The cross effects just derived are the off-diagonal entries of the equilibrium system's Jacobian, up to sign. They need not be nonzero for every experiment and signal pair; nevertheless their presence prevents a generic signal-by-signal optimization of verification.

\begin{proposition}[Local response of equilibrium verification to disclosure]
\label{prop:local-disclosure-verification-coupling}
Suppose $\mathcal B$ is continuously differentiable near an interior equilibrium $(e_0,t_0)$ and $J=D_e\mathcal B(e_0,t_0)$ is nonsingular. Then the equilibrium branch is locally differentiable and satisfies
\begin{equation}\label{eq:disclosure-effort-response}
 \frac{de^*(t_0)}{dt}=-J^{-1}\mathcal B_t(e_0,t_0).
\end{equation}
Thus the equilibrium effort vector has a nonzero first-order response along the disclosure family if and only if $\mathcal B_t(e_0,t_0)\neq0$. This local condition is not an assertion that every equilibrium or every optimal disclosure policy is nonseparable.
\end{proposition}
\begin{proof}
Apply the implicit-function theorem to $\mathcal B(e,t)=0$. The equivalence follows from invertibility of $J$.
\end{proof}

The first proposition establishes exactly when another verification margin enters an agent's private return to information acquisition. The second shows how a feasible change in disclosure propagates through the whole verification system. These are distinct effects: the former holds at a fixed experiment, whereas the latter concerns equilibrium readjustment to a change in the experiment. Their economic consequence is that the seller's revenue comparison must be computed using the induced equilibrium $G^{p,*}(\pi)$ rather than holding verification behavior fixed. The revenue identity of Subsection~\ref{subsec:two-tradeoffs} and the global implementability criterion of Subsection~\ref{subsec:equilibrium-implementability} already incorporate this dependence. In contrast, separated-support policies can eliminate certain cross-signal effects, as demonstrated by the restricted-class results in Section~6.

This establishes a precise channel through which the benchmark disclosure structure---complete revelation in a lower valuation region and pooling in a higher region---can lose its optimality when verification is available. The channel explains the possibility; the strict revenue comparisons of Section~4 establish actual failures. Section~6 develops tests for other candidate disclosure structures without requiring a closed-form solution to the unrestricted design problem.

\section{Tractable Optimality Conditions and Comparative Statics}
\label{sec:tractable-conditions}

The general results of Section~5 establish what it means for a bid distribution to be equilibrium implementable and globally revenue maximizing. Their unrestricted formulation, however, need not yield an explicit optimal disclosure policy, even for simple valuation distributions. We therefore turn to the operational consequences of the characterization: necessary conditions from feasible changes in disclosure; posterior contact and extreme-point methods that simplify auxiliary problems; exact structural reductions within specified policy classes; and comparative statics of the resulting conditions. The aim is not a universal closed-form solution or an unrestricted finite-support theorem. Rather, the goal is to formulate correct restrictions that candidate policies must satisfy and exploit their violation to exclude policies from global optimality.

The convex-order approach in Bergemann et al. (2022) and the extreme-point geometry of Kleiner, Moldovanu, and Strack (2021) and Kleiner, Moldovanu, Strack, and Whitmeyer (2026) motivate these reductions. Because verification is endogenous, an extreme-point property of a \emph{fixed-$H$ auxiliary problem} does not by itself establish an extreme-point optimum for the seller. We distinguish exact identities, conditional structural restrictions, and research extensions accordingly.

\subsection{Equilibrium-Adjusted Necessary Conditions and Posterior Reductions}
\label{subsec:extreme-point-conditions}

The exact global characterization in Subsection~\ref{subsec:equilibrium-implementability} does not presuppose finitely many posterior signals, yet direct maximization over all experiments remains demanding. Extreme-point and contact-set methods provide a disciplined route to simplification. In the classical problem, convex order and the geometry of mean-preserving contractions can make extreme experiments particularly useful. Kleiner, Moldovanu, and Strack (2021) characterize extreme points under majorization restrictions, while Kleiner, Moldovanu, Strack, and Whitmeyer (2026) study extreme mean-preserving contractions and the geometry of their finite supports. Here the additional verification-equilibrium fixed point changes the seller's objective, so a reduction valid for a linearized fixed-distribution auxiliary problem must not be silently promoted to an unrestricted finite-signal representation.

For fixed $G^p$ and $g$, Proposition~\ref{prop:conditional-posterior-contact} is already an exact support restriction whenever a minimizing posterior price exists. In particular, a finite contact set reduces the corresponding auxiliary problem to finitely many posterior weights. This does not yet prove that the seller's globally optimal disclosure can be selected from that contact set, because the equilibrium bid distribution must be recomputed when the experiment changes. We therefore separate conditional algebraic reductions from general necessary optimality tests.

\subsubsection{Necessary conditions from feasible disclosure changes}

A candidate seller optimum must resist every admissible improvement, including perturbations that alter incumbent posteriors or introduce previously absent signals. For a regular family of experiments these conditions can be expressed in terms of the equilibrium response of verification and the induced change in the revenue functional. They are useful even when no global explicit optimum is available. In particular, a strictly positive one-sided equilibrium-adjusted revenue change rules out that candidate. The following formulas record regularity-qualified tools for such tests; the moving-atom cases require exact boundary treatment, as explained below.
\noindent\textbf{Verification effects holding other efforts fixed.}
For $N\ge2$, define $\omega_N(x)=x^{N-2}(1-x)$. When $N\ge3$, this weight increases below $g^\dagger=(N-2)/(N-1)$ and decreases above it. For $N=2$, $\omega_2(x)=1-x$ is nonincreasing throughout $[0,1]$. The weight summarizes how a change in a bidder's final valuation distribution affects the expected second-highest bid.
A seller's direct marginal incentive to encourage verification depends on where the posterior distribution is located relative to rival bids. The next proposition identifies the sign when the rank-dependent revenue weight is monotone on the relevant value interval.
\begin{proposition}[Private verification's partial revenue effect]\label{sign}
Holding the remaining effort coordinates fixed,
\begin{equation}\label{B} B_s:=\frac{\partial R_N(G^{p}_e)}{\partial e_s}=-N(N-1)\pi_s\int_0^M \omega_N(G^{p}_e(b))\Delta_s(b)db.\end{equation}
If $b\mapsto \omega_N(G^{p}_e(b))$ is nondecreasing on the convex hull of $\mathrm{supp}(\mu_s)\cup\{m_s\}$, then $B_s\ge0$. If that function is nonincreasing on the same hull, then $B_s\le0$. For $N\ge3$, monotonicity follows when $G^{p}_e(b)\le g^\dagger$ throughout the hull (positive case), or $G^{p}_e(b)\ge g^\dagger$ throughout (negative case). If the relevant distribution interval crosses the peak, no general sign follows. For $N=2$, the weight is nonincreasing everywhere, so $B_s\le0$ for every posterior and every fixed profile of the remaining efforts.
\end{proposition}
\begin{proof} Differentiating the seller-revenue formula in Section~2 gives \eqref{B}. Since $\int\Delta_s(b)db=0$ and $\Delta_s$ is nonnegative below $m_s$ and nonpositive above $m_s$, the integral of a nondecreasing weight times $\Delta_s$ is nonpositive; the opposite applies to a nonincreasing weight. The piecewise monotonicity of $\omega_N$ gives the last statement.\end{proof}
\begin{remark}Positive definiteness of $\mathrm{diag}(\pi)J$ does \emph{not} make $J^{-1}$ entrywise positive. Therefore separate partial revenue signs never, by themselves, determine the sign of the total policy effect. In particular, for generic overlapping conditional supports the cross-signs in $J$ are ambiguous.\end{remark}

\noindent\textbf{Experiment perturbations with an unchanged signal alphabet.}
Verification effects computed with other efforts held fixed are not seller-policy effects. An experiment variation also changes the posterior laws and equilibrium verification. For a regular finite-dimensional path of Bayes-plausible experiments, write $\mathcal B_s(e,t)=c'(e_s)-A_s(G^{p}_{e,t})$ and $J=D_e\mathcal B$. The next formula keeps these two sources of change separate.
For a \emph{fixed finite signal alphabet} and differentiable feasible path $t\mapsto(\pi_s(t),\mu_s(t))$ that exactly preserves $F$, suppose the symmetric equilibrium is unique, interior, the Jacobian $J=D_e\mathcal B$ is nonsingular, and the bid cdf has an admissible directional derivative in the $L^1$ sense. Write $\mathcal B_s(e,t)=c'(e_s)-A_s(G^{p}_{e,t})$. All derivatives in the following two displays are evaluated at the reference experiment $t=0$ and its interior equilibrium $e^*=e(0)$: $\mathcal B_t=\partial_t\mathcal B(e^*,0)$, $G_t^p=\partial_tG^p_{e^*,t}|_{t=0}$, and $D_eG^p$ is the derivative with respect to the incumbent effort vector at $(e^*,0)$. The dot denotes the total derivative along the equilibrium path, whereas $G_t^p$ holds incumbent efforts fixed. The chain rule and implicit-function theorem (see \citealp{Rudin1976}) give
\begin{equation}\label{imp} e'(0)=-J^{-1}\mathcal B_t,\qquad \dot G^p=G^{p}_t-D_eG^p\,J^{-1}\mathcal B_t,\end{equation}
\begin{equation}\label{var} \boxed{\dot R=-N(N-1)\int_0^M G^p(b)^{N-2}(1-G^p(b))\left[G^{p}_t(b)-D_eG^p(b)J^{-1}\mathcal B_t\right]db.}\end{equation}
If the candidate experiment is a local seller optimum, $\dot R\le0$ for every admissible \emph{one-sided} path. This is a necessary condition, not a sufficient global certificate. When posterior bid atoms move, ordinary pointwise cdf differentiation is \emph{not} justified at their locations; \eqref{var} remains valid when the indicated cdf derivatives converge in $L^1$ or by directly differentiating piecewise integrals with moving boundaries. Finite-signal paths do not exhaust experiments that \emph{introduce new signals}.

\noindent\textbf{Rare-signal splitting and boundary directions.}
The experiment space allows the seller to introduce a signal that has zero probability at the original policy. Such a variation cannot be treated simply as a differentiable perturbation of fixed signal weights. It is particularly useful for detecting failure of local optimality within an apparently well-behaved cutoff family. We extract a small subposterior $\nu$ from a pooled posterior $\mu$, adjust the residual posterior exactly to preserve the prior, and compute the entrant's limiting best response separately from the incumbent's effort adjustment.
Let an incumbent signal have prior mass $\pi_0$, posterior $\mu$ and incumbent effort $e$. Let $\nu$ be a probability measure absolutely continuous with respect to $\mu$, with essentially bounded $d\nu/d\mu$. Split fraction $\varepsilon$ into a new signal of posterior $\nu$, leaving posterior $\mu_{0,\varepsilon}=(\mu-\varepsilon\nu)/(1-\varepsilon)$. For $\varepsilon$ small enough this is a valid measure, and Bayes plausibility is preserved \emph{exactly}. Write $m=\E_\mu v$, $m_\nu=\E_\nu v$ and let $r$ be the new type's limiting best response against the original equilibrium bid cdf $G^{p,*}$. For any $C^1$ test function $\varphi$, the fixed-incumbent-effort weak directional derivative $\eta^{\mathrm{dir}}$ of the bid law (acting on smooth test functions, and not necessarily a finite signed measure) satisfies
\begin{equation}\label{measure}\boxed{\int\varphi\,d\eta^{\rm dir}=\pi_0\left[(1-e)\{\varphi(m_\nu)-\varphi(m)-(m_\nu-m)\varphi'(m)\}+(r-e)\{\E_\nu\varphi(v)-\varphi(m_\nu)\}\right].}\end{equation}
The identity \eqref{measure} remains valid for a moving incumbent mean, but the derivative then contains a derivative of a Dirac mass and is not an ordinary finite signed measure. To obtain a direct integral test, restrict temporarily to \emph{mean-preserving splits} satisfying $m_\nu=m$. In this case the incumbent mean does not move, and \eqref{measure} reduces to the finite signed measure $\eta^{\mathrm{dir}}=\pi_0(r-e)(\nu-\delta_m)$. For incumbent signals indexed by $t$, let $b_t=\partial_\varepsilon\mathcal B_t(e^*,0;r)$ include \emph{both} the induced bid cdf and the changing incumbent posterior, let $J$ be the incumbent effort Jacobian, and define $\eta_t=\pi_t(\mu_t-\delta_{m_t})$. Assume the incumbent equilibrium is interior, differentiable along the split, and has nonsingular $J$. Then the full bid-measure derivative is the finite signed measure $\eta^{\rm dir}-\sum_t\eta_t[J^{-1}b]_t$. Its cdf derivative is bounded and therefore $L^1$-admissible, so the seller's rare-signal derivative is
\begin{equation}\label{split}\D_{s}(\nu)=-N(N-1)\int_0^M G^{p,*}(x)^{N-2}(1-G^{p,*}(x))\left[\eta^{\rm dir}([0,x])-\sum_t\eta_t([0,x])[J^{-1}b]_t\right]dx.\end{equation}
At a local optimum, $\D_s(\nu)\le0$ for every admissible mean-preserving split satisfying these regularity conditions. A strictly positive value disproves local optimality. For $m_\nu\ne m$, the weak derivative \eqref{measure} is still exact, but \eqref{split} is \emph{not} asserted: moving incumbent atoms must instead be evaluated by differentiating the original revenue integral with its moving boundaries, including the boundary contribution of the atom. In particular, a distributional derivative of an atom cannot be inserted directly into \eqref{split} as though it were the cdf of a finite signed measure. This distinction is essential for a valid one-sided test.

\noindent\textbf{A fixed-$H$ posterior-splitting identity.}
The local methods can also be organized around an exact algebraic identity. Fix a candidate bid law $H$ and quadratic verification costs $c_\kappa(e)=\kappa e^2/2$. Let $\nu$ be a finite signed Borel measure and $\mu^\pm=\mu\pm\varepsilon\nu$ be probability measures with the same mean as $\mu$, where $\nu(V)=\int v\,\nu(dv)=0$, and suppose the three verification best responses against this \emph{fixed} $H$ remain interior. Then $e(\mu^\pm,H)=e(\mu,H)\pm (\varepsilon/\kappa)\int\Phi_H\,d\nu$, so a direct expansion yields
\begin{equation}\label{eq:fixed-H-splitting}
\frac{T_H(\mu^+)+T_H(\mu^-)}{2}-T_H(\mu)
=\frac{\varepsilon^2}{\kappa}\left(\int_V\Phi_H\,d\nu\right)\nu.
\end{equation}
The identity displays an endogenous-verification term that is absent from a purely linear posterior-mean comparison. It is an equality for a \emph{fixed rival-bid law}; after the experiment changes, the equilibrium bid law generally changes too. In particular, \eqref{eq:fixed-H-splitting} alone does not establish the seller's equilibrium-adjusted sign condition or a universal finite-support theorem. Establishing the applicability of any further contact, curvature, or few-point restrictions requires separate hypotheses on the relevant equilibrium-adjusted marginal functional. Such restrictions are retained as conditional research directions, not asserted here as unrestricted optimality theorems.

\subsection{Structural Characterization under Separated Posterior Supports}
\label{subsec:separated-interval-pooling}

The unrestricted implementability conditions in the preceding implementability analysis account for Bayes plausibility and verification incentives but do not prescribe the shape of an optimal experiment. To see how the two trade-offs identified in the revenue decomposition above operate in a tractable disclosure design, we consider policies that pool valuations within disjoint intervals and reveal valuations outside them. This is a genuine restriction on the seller's experiment, not a claim about unrestricted optimality. Because a bidder's final bid stays within the interval associated with the observed signal, verification cannot reverse the ordering of bids from disjoint intervals. The resulting separation permits explicit verification and revenue formulas, followed by a global optimization within the interval-pooling class.

Fix a finite collection of pairwise disjoint intervals $I_j=[a_j,b_j]\subseteq V=[0,M]$, $j=1,\ldots,J$, with $p_j=F(b_j)-F(a_j)>0$. Each interval receives a distinct pooled signal, while valuations outside the union of these intervals are fully disclosed. Conditional on pooled signal $j$, the posterior $\mu_j$ is the prior restricted to $I_j$ and normalized by $p_j$; its mean is
\[
m_j=\frac{1}{p_j}\int_{a_j}^{b_j}v\,F(dv).
\]
The bidder initially bids $m_j$ after pooled signal $j$, and bids $v_i\in I_j$ if verification succeeds. In the analysis below $F$ is absolutely continuous, so interval endpoint conventions do not affect these formulas.

\begin{definition}[Finite Interval-Pooling Disclosure Experiment]
\label{def:finite-interval-pooling}
A disclosure experiment $\pi\in\Pi_F$ is a \emph{finite interval-pooling experiment} if, for some finite pairwise disjoint family $(I_j)_{j=1}^J$ as above, it pools exactly the values in each $I_j$ and fully reveals every value outside $\bigcup_{j=1}^J I_j$.
\end{definition}

\begin{proposition}[Exact separability of interval pooling]
\label{prop:interval-separability}
Suppose $N\geq2$, $F$ is absolutely continuous on $[0,M]$, and
$c:[0,1]\to\mathbb R$ is continuously differentiable and strictly
convex, as in Section 3. Under any finite interval-pooling experiment
of the preceding form, the following conclusions hold.

\begin{enumerate}
\item For $x\in[a_j,b_j]$, the symmetric final-bid distribution is
\[
G^p(x)=
\begin{cases}
 F(a_j)+e_j\bigl(F(x)-F(a_j)\bigr),&a_j\leq x<m_j,\\
 F(b_j)-e_j\bigl(F(b_j)-F(x)\bigr),&m_j\leq x\leq b_j.
\end{cases}
\]
Outside the pooling intervals, $G^p(x)=F(x)$.

\item Let
\[
\Delta_j(x)=\frac{F(x)-F(a_j)}{p_j}
            -\mathbf 1\{m_j\leq x\},\qquad a_j\leq x\leq b_j.
\]
The acquisition value following signal $j$ is
\[
 A_j(e_j)=-\int_{a_j}^{b_j}[G^p(x)]^{N-1}\Delta_j(x)\,dx,
\]
and it depends on no other effort coordinate. Moreover,
\[
 \frac{dA_j(e_j)}{de_j}
 =-(N-1)p_j\int_{a_j}^{b_j}
 [G^p(x)]^{N-2}\Delta_j(x)^2\,dx\leq0.
\]
The unique symmetric equilibrium effort $e_j^*$ is characterized by
\[
\begin{cases}
 e_j^*=0,&A_j(0)\leq c'(0),\\
 c'(e_j^*)=A_j(e_j^*),&0<e_j^*<1,\\
 e_j^*=1,&A_j(1)\geq c'(1).
\end{cases}
\]
The inequalities describe the respective boundary cases; otherwise
the interior equation determines the effort uniquely.

\item With
\[
 \Psi_N(u)=1-Nu^{N-1}+(N-1)u^N,
\]
the seller's revenue satisfies the exact decomposition
\[
 R_N(G^p)=R_N(F)+\sum_{j=1}^J
 \int_{a_j}^{b_j}
 \bigl[\Psi_N(G^p(x))-\Psi_N(F(x))\bigr]dx.
\]
Each summand depends only on its own pooling interval and its
associated equilibrium effort.
\end{enumerate}
\end{proposition}

\begin{proof}
Fix an interval $[a_j,b_j]$. Its signal has prior probability $p_j$.
Bids generated by this signal remain in $[a_j,b_j]$, while bids generated
by signals for intervals entirely below or above it remain on the
corresponding side. Fully disclosed bids outside the pooling intervals
also preserve their original values. Therefore the probability of a bid
below $a_j$ is $F(a_j)$, and that of a bid at most $b_j$ is $F(b_j)$,
regardless of verification efforts elsewhere. Conditional on signal $j$,
the final bid equals $m_j$ with probability $1-e_j$ and has distribution
$\mu_j$ with probability $e_j$. These observations give the two displayed
cdf formulas. They also imply $G^p=F$ outside the pooling intervals.

The acquisition-value identity in Section 3 gives
$A_j=-\int[G^p(x)]^{N-1}\Delta_j(x)\,dx$.
The integrand vanishes outside $[a_j,b_j]$. Within that interval,
$\partial G^p(x)/\partial e_j=p_j\Delta_j(x)$ at every point other than
the posterior mean, which is immaterial for integration. Differentiation
under the integral establishes the displayed nonpositive derivative.
Thus $A_j(e_j)$ is nonincreasing in its own effort and is independent
of all other effort coordinates. Since $c'$ is strictly increasing,
the best-response conditions have a unique solution in $[0,1]$ for
each signal. All other signals are fully revealing and therefore
induce zero effort. This proves parts (1) and (2).

Finally, expected revenue is the integral of $\Psi_N(G^p)$ over $V$.
Subtract the full-disclosure revenue $R_N(F)$ and use the equality
$G^p=F$ outside the disjoint pooling intervals. The remaining integrals
have disjoint domains, yielding part (3).
\end{proof}

The proposition applies to general absolutely continuous priors and
any number of bidders. It does not rely on the uniform distribution or
on quadratic costs. Its limitation is equally important: experiments
with overlapping posterior supports need not have this separability,
and the unrestricted seller problem cannot be reduced to independent
pooling intervals without an additional optimality argument.

For the three-bidder uniform benchmark, the decomposition has a closed
form. It is convenient to write $c(e)=\kappa e^2/2$, where $\kappa>0$.

\begin{corollary}[Closed-form verification and revenue]
\label{cor:uniform-interval-formulas}
Suppose $N=3$, $F$ is uniform on $[0,1]$, and
$c(e)=\kappa e^2/2$. For a pooling interval $[a,b]$, set
$d=b-a$ and $m=(a+b)/2$. Its equilibrium effort is
\[
 e_\kappa(a,b)
 =\min\left\{1,
 \frac{3md^2}{12\kappa+2md^2}\right\}.
\]
The interval's equilibrium revenue contribution relative to full
disclosure is
\[
 W_\kappa(a,b)
 =\frac{(a+b-1)(b-a)^3}{4}
 \bigl(1-e_\kappa(a,b)\bigr)
 \bigl(2-e_\kappa(a,b)\bigr).
\]
Consequently, any finite collection of disjoint pooling intervals
yields
\[
 R_3(G^p)=\frac12+\sum_{j=1}^J W_\kappa(a_j,b_j).
\]
In this policy class, a pooling interval contributes strictly positive
revenue if and only if $a+b>1$ and $e_\kappa(a,b)<1$.
\end{corollary}

\begin{proof}
Write $x=a+y$, where $0\leq y\leq d$. By Proposition
\ref{prop:interval-separability}, the final-bid cdf on the interval is
\[
 G^p(a+y)=
 \begin{cases}
 a+ey,&0\leq y<d/2,\\
 a+d-e(d-y),&d/2\leq y\leq d.
 \end{cases}
\]
The posterior cdf is $y/d$, so the acquisition value is
\[
\begin{aligned}
 A(e)
 ={}&-\int_0^{d/2}(a+ey)^2\frac{y}{d}\,dy\\
 &-\int_{d/2}^d
  \bigl(a+d-e(d-y)\bigr)^2\frac{y-d}{d}\,dy\\
 ={}&\frac{md^2}{12}(3-2e).
\end{aligned}
\]
The interior equilibrium condition $\kappa e=A(e)$ gives
$e=3md^2/(12\kappa+2md^2)$; imposing the upper effort bound
gives the stated formula. Since the expression for $A(e)$ is
strictly decreasing whenever $d>0$ and $m>0$, the equilibrium is
unique.

For three bidders, $\Psi_3(u)=1-3u^2+2u^3$. Thus the revenue
contribution is
\[
\begin{aligned}
 W_\kappa(a,b)
 ={}&\int_0^{d/2}
 \bigl[\Psi_3(a+ey)-\Psi_3(a+y)\bigr]dy\\
 &+\int_{d/2}^d
 \bigl[\Psi_3(a+d-e(d-y))-\Psi_3(a+y)\bigr]dy.
\end{aligned}
\]
Integrating the two cubic polynomials and substituting the
equilibrium effort gives
\[
 W_\kappa(a,b)
 =\frac{(a+b-1)d^3}{4}(1-e)(2-e).
\]
Full disclosure yields revenue $1/2$, and the additivity assertion
follows from Proposition \ref{prop:interval-separability}.
The sign characterization follows from $0\leq e\leq1$.
\end{proof}

A useful consequence is that the seller never benefits, within this
policy class, from pooling an interval whose midpoint lies below
$1/2$. More surprisingly, positive pooling opportunities exist
arbitrarily close to every point above $1/2$: for any $x>1/2$,
a sufficiently short interval around $x$ has
$e_\kappa(a,b)<1$ and $W_\kappa(a,b)>0$.
This observation restricts how an optimal finite interval partition
can be arranged.

\begin{proposition}[Contiguous upper pooling and cutoff conditions]
\label{prop:contiguous-upper-pooling}
Maintain the hypotheses of Corollary
\ref{cor:uniform-interval-formulas}. If a revenue-maximizing policy
within the class of finite interval-pooling experiments is attained,
then there exists a revenue-maximizing representative, obtained by
removing nonpositive-contribution pools if necessary, whose pooling intervals
form a contiguous partition
\[
 [q,x_1],\ [x_1,x_2],\ \ldots,\ [x_{J-1},1]
\]
for some $q\leq1/2$, and every retained interval has strictly positive revenue contribution.
Valuations below $q$ are fully disclosed.

At an interior boundary $x_j$ where the adjacent equilibrium efforts
are on differentiable branches, a necessary cutoff condition is
\[
 \partial_2 W_\kappa(x_{j-1},x_j)
 +\partial_1 W_\kappa(x_j,x_{j+1})=0.
\]
If $q\in(0,1/2)$ is interior and the first interval is regular,
its lower boundary satisfies
\[
 \partial_1 W_\kappa(q,x_1)=0.
\]
These conditions are necessary, not sufficient, for an optimal
partition.
\end{proposition}

\begin{proof}
Deleting any interval with a negative contribution strictly raises
revenue without affecting efforts in the remaining intervals.
A zero-contribution interval can likewise be replaced by full
disclosure without affecting revenue or other efforts. If this
creates an open region above $1/2$, introducing a sufficiently
small pooling interval there strictly increases revenue.
Thus any maximizing policy can be replaced, without loss, by a
representative containing only positive-contribution pools.

Fix such a representative. Suppose an open subinterval of $(1/2,1)$
is fully disclosed.
Choose a smaller interval $[a,b]$ within it. Its midpoint is
above $1/2$, and by taking $b-a$ sufficiently small we obtain
$e_\kappa(a,b)<1$. Corollary
\ref{cor:uniform-interval-formulas} then gives
$W_\kappa(a,b)>0$. Adding this new signal leaves all existing
efforts unchanged and strictly improves revenue, a contradiction.
Hence no open part of $(1/2,1)$ can be left outside the union
of pools.

Every positive-contribution pool has midpoint above $1/2$;
therefore at most one such pool can start below $1/2$, and
it necessarily crosses $1/2$. Because there are finitely many
intervals and no open gap above $1/2$, the intervals must be
adjacent, their last right endpoint must be $1$, and their first
left endpoint $q$ cannot exceed $1/2$. This proves the first
assertion.

At a differentiable interior boundary, varying that boundary
changes only the two adjacent interval contributions.
Differentiating their sum gives the displayed necessary condition.
The condition for an interior initial cutoff follows analogously.
\end{proof}

This result retains the economic distinction between revelation of
low values and pooling of high values found in BHMSW. Within the
interval-pooling class, however, the upper region may contain
several adjacent pools with different posterior means and different
endogenous verification efforts. The number and boundaries of these
pools are part of the seller's optimization problem.

We can characterize the global optimum over this restricted class
without assuming a fixed number of pools. Let $\mathfrak I(x)$ be
the family of all finite collections of pairwise disjoint intervals
contained in $[0,x]$, including the empty collection, and define
\[
 D_\kappa(x)
 =\sup_{\mathcal I\in\mathfrak I(x)}
   \sum_{[a,b]\in\mathcal I}W_\kappa(a,b).
\]
The supremum is finite because
$|W_\kappa(a,b)|\leq(b-a)^3/2$ and the interval lengths sum
to at most one.

\begin{theorem}[Bellman and dual characterization of interval pooling]
\label{thm:interval-pooling-bellman}
Under the hypotheses of Corollary
\ref{cor:uniform-interval-formulas}, the maximum revenue
supremum over all finite interval-pooling experiments is
\[
 R_{\mathrm{IP}}^*=\frac12+D_\kappa(1).
\]
The value function satisfies, for every $x\in[0,1]$,
\[
 D_\kappa(x)
 =\max\left\{0,
 \sup_{0\leq a<b\leq x}
 \bigl[D_\kappa(a)+W_\kappa(a,b)\bigr]
 \right\}.
\]
Equivalently,
\[
 D_\kappa(1)=\inf_U U(1),
\]
where the infimum is over nondecreasing functions
$U:[0,1]\to\mathbb R$ satisfying
\[
 U(0)=0,\qquad
 U(b)-U(a)\geq W_\kappa(a,b)
 \quad\text{for all }0\leq a<b\leq1.
\]
Thus a feasible $U$ whose terminal value equals the revenue gain
of a particular interval partition certifies that partition's
global optimality within the interval-pooling class.

For a finite grid $0=x_0<x_1<\cdots<x_K=1$, the exact restricted
optimum is computed from $D_0=0$ and
\[
 D_j=\max\left\{
 D_{j-1},\quad
 \max_{0\leq i<j}
 \bigl[D_i+W_\kappa(x_i,x_j)\bigr]
 \right\},\qquad j=1,\ldots,K.
\]
Its revenue is $1/2+D_K$. For nested uniform grids with mesh tending to zero, the grid revenues converge to $R_{\mathrm{IP}}^*$; this does not assert existence of an optimizing finite partition in the unrestricted finite-interval class.
\end{theorem}

\begin{proof}
The revenue identity follows from Corollary
\ref{cor:uniform-interval-formulas}. For the Bellman identity,
consider the last interval $[a,b]$ of a nonempty finite
collection contained in $[0,x]$. Every earlier interval lies
in $[0,a]$, so its total contribution is at most
$D_\kappa(a)+W_\kappa(a,b)$. Conversely, given $a<b\leq x$,
a finite collection in $[0,a]$ can approximate $D_\kappa(a)$
arbitrarily closely, and appending $[a,b]$ preserves feasibility.
The empty collection contributes zero. Taking suprema proves
the identity.

If $U$ satisfies the displayed inequalities, then for any ordered
collection $[a_1,b_1],\ldots,[a_J,b_J]$,
\[
 \sum_{j=1}^JW_\kappa(a_j,b_j)
 \leq\sum_{j=1}^J[U(b_j)-U(a_j)]\leq U(1),
\]
where the second inequality uses monotonicity and disjointness.
Hence $D_\kappa(1)\leq U(1)$ for every feasible $U$.
On the other hand, the definition of $D_\kappa$ makes it
nondecreasing, with $D_\kappa(0)=0$, and appending a single
interval to a nearly optimal collection in $[0,a]$ gives
\[
 D_\kappa(b)-D_\kappa(a)\geq W_\kappa(a,b).
\]
Thus $U=D_\kappa$ is feasible and achieves equality in the
dual bound.

For the grid recursion, either no pooled interval ends at $x_j$,
in which case the optimal gain is $D_{j-1}$, or the last interval
is $[x_i,x_j]$, in which case earlier intervals contribute at
most $D_i$. Maximizing over these alternatives proves the
recursion. Finally, $W_\kappa(a,b)$ is continuous in its
endpoints, including at the boundary between interior effort and $e=1$. Each fixed finite collection can therefore be
approximated by a grid-aligned collection with arbitrarily
small revenue loss as the grid becomes fine. Taking the
supremum over finite collections proves convergence.
\end{proof}

The Bellman and dual characterizations are global within the
interval-pooling class. They do not prove that a revenue-maximizing
experiment in the unrestricted space $\Pi_F$ has disjoint posterior
supports or finite support. Establishing such a reduction would
require an additional unrestricted optimality argument, not merely
the equilibrium separability proved here.

\subsection{Comparative Statics and Optimality Tests}
\label{subsec:optimal-disclosure-comparative-statics}

A characterization is most valuable when it can be used to study how optimal policies respond to changes in economic primitives. The fixed-experiment result in Proposition~\ref{prop:verification-cost-comparative-statics} is a first step: it establishes an aggregate verification response to the cost coefficient, but does not say how the seller reoptimizes disclosure. The present subsection explains how to carry the optimization conditions from Section~5 and Subsection~\ref{subsec:extreme-point-conditions} into comparative statics without requiring an explicit optimal policy.

\noindent\textbf{General parameter stability.}
Let $\lambda$ vary in a parameter space, affecting the verification cost or another primitive while leaving the admissible experiment space fixed. Write $G^{p,*}(\pi;\lambda)$ for the unique equilibrium final-bid distribution and define
\[
 \mathcal V(\lambda)=\max_{\pi\in\Pi_F}R_N(G^{p,*}(\pi;\lambda)),\qquad
 \Pi^*(\lambda)=\arg\max_{\pi\in\Pi_F}R_N(G^{p,*}(\pi;\lambda)).
\]
The existence theorem in Section~3 applies pointwise in $\lambda$. To compare optimizers across parameters, continuity must hold jointly in $(\pi,\lambda)$, rather than merely for each fixed parameter. This is a separate regularity requirement, to be verified for the particular parameter family.

\begin{proposition}[Stability of optimal disclosure under joint equilibrium continuity]
\label{prop:optimal-disclosure-stability}
Suppose the parameter space is metric, $\Pi_F$ is endowed with its compact weak topology, and $(\pi,\lambda)\mapsto G^{p,*}(\pi;\lambda)$ is jointly weakly continuous. Then $\mathcal V$ is continuous and $\Pi^*$ is a nonempty compact-valued, upper-hemicontinuous correspondence.
\end{proposition}
\begin{proof}
The seller-revenue functional $R_N$ is continuous under weak convergence of bid distributions on the compact interval $V$: expected second-highest bids are expectations of a bounded continuous function of $N$ independent bids. Hence $(\pi,\lambda)\mapsto R_N(G^{p,*}(\pi;\lambda))$ is jointly continuous. Compactness of $\Pi_F$ and Berge's maximum theorem (\citealp{Berge1963}) yield the conclusions.
\end{proof}

The proposition deliberately imposes joint equilibrium continuity rather than inferring it from the pointwise convex-potential argument alone. Its hypotheses can be checked in regular parametric settings; the result does not imply that an optimizer is unique or that its posterior support changes continuously when the optimum switches between distinct experiments.

\noindent\textbf{Local changes within a reduced policy class.}
Suppose a valid structural reduction represents a policy and its associated equilibrium by a finite vector $z$, and a locally stable set of binding optimality and equilibrium conditions by $\mathcal E(z,\lambda)=0$, with the remaining inequalities strict. This representation may arise, for example, from finitely many conditional posterior contacts combined with a finite parametrization of the relevant posteriors. Merely asserting finitely many signals is insufficient if each posterior remains a free probability measure.

\begin{proposition}[Local comparative statics of a regular reduced system]
\label{prop:reduced-local-comparative-statics}
Let $\mathcal E:\mathbb R^d\times\mathbb R\to\mathbb R^d$ be continuously differentiable near $(z_0,\lambda_0)$, with $\mathcal E(z_0,\lambda_0)=0$ and nonsingular $D_z\mathcal E(z_0,\lambda_0)$. Then there exists a unique continuously differentiable local solution branch $z$ with $z(\lambda_0)=z_0$ and
\[
 \frac{dz}{d\lambda}(\lambda_0)
 =-[D_z\mathcal E(z_0,\lambda_0)]^{-1}D_\lambda\mathcal E(z_0,\lambda_0).
\]
If $z_0$ describes an optimal disclosure, the formula describes the locally continued optimal policy only when feasibility, the active set, and optimality against competing candidates remain valid along the branch.
\end{proposition}
\begin{proof}
The implicit function theorem (see \citealp{Rudin1976}) yields the branch and derivative. The last assertion is a scope condition: solving the active equations does not itself establish global seller optimality.
\end{proof}

These conditions also clarify how to locate possible structural transitions. A posterior weight may vanish, a previously slack inequality may bind, or two globally feasible policy branches may deliver equal revenue. Such events provide candidate parameter thresholds, not an unconditional theorem that a specific pooling region must expand or contract with verification costs. Changes in the discrete bidder count $N$ require comparisons across distinct models rather than differentiation in $N$.

An important application is to test the benchmark's disclosure structure. If a cost or distribution parameter makes one of the valid necessary optimality inequalities fail, the corresponding candidate policy is excluded without first identifying the seller's unrestricted optimal experiment. Section~4 provides direct revenue comparisons; reduced inequalities can potentially give sharper parameter-dependent exclusion tests.

\section{Conclusion}

This paper revisits optimal information disclosure in auctions through the familiar principle of ``trust, but verify.'' The fundamental insight is that information disclosure determines not only what bidders learn from the seller, but also what they have incentives to learn on their own. This creates two interrelated trade-offs. The seller must balance the direct effects of disclosure on competition and information rents against its indirect effects on bidders' incentives to acquire information privately. Each bidder, in turn, must weigh the benefits of more accurate information against the costs of verification, taking into account the information-acquisition decisions of competing bidders. These trade-offs interact through equilibrium: withholding information may preserve the seller's informational advantage, but it may also induce bidders to reduce that advantage through private verification. The seller controls initial disclosure, but not the information ultimately acquired by bidders. In the spirit of Hurwicz's incentive-based perspective and the structural-invariance concern highlighted by the Lucas critique, the policy-induced verification response cannot be treated as fixed when disclosure changes. This fundamental tension explains why the distribution-independent optimal disclosure result of Bergemann et al. (2022)---revealing low values while pooling high values at a critical quantile determined solely by the number of bidders---is no longer generally revenue-maximizing even for the familiar uniform and binary valuation distributions, once private verification is endogenous. Thus, the robustness of the benchmark policy across valuation distributions does not extend generally to an environment in which bidders can respond to disclosure by acquiring information for themselves.

Recognizing the limitations of the benchmark policy leads to a more general question: how should a seller design information disclosure when bidders' subsequent information acquisition is determined endogenously in equilibrium? To address this question, we develop a framework that connects bidders' Bayesian Nash verification equilibrium with the seller's optimal information design problem. We establish existence and uniqueness of the symmetric verification equilibrium and the existence of an optimal disclosure policy, characterize equilibrium implementability, and derive an exact revenue decomposition that identifies the competing effects of disclosure and verification. Building on these results, we provide a general characterization of global optimality, together with tractable necessary conditions, structural reductions, and comparative-static results under the relevant conditions. The resulting analysis shows why optimal disclosure cannot generally be determined by treating information acquisition as fixed or by optimizing the final distribution of bidders' information independently of the incentives that generate it. The relevant object of design is the interaction between the seller's information policy and the equilibrium information-acquisition decisions it induces.

The economic mechanism extends beyond the particular auction environment studied here. Reserve prices and alternative auction formats, including uniform-price auctions, introduce additional margins through which disclosure may influence private verification and seller revenue. Similarly, allowing costly participation and private verification to be jointly endogenous would connect information acquisition with bidders' entry decisions. Establishing optimal disclosure in these environments requires further analysis of the corresponding equilibria. Nevertheless, the central lesson is more general: when agents can privately produce information in response to what is publicly disclosed, information design and information acquisition become inseparable parts of the same economic problem. Optimal disclosure must therefore account not only for the information the designer provides, but also for the information that agents are induced to acquire for themselves.

\end{document}